\documentclass[structabstract]{aa} 
\usepackage{graphicx} 
\usepackage{float}

\usepackage{txfonts}
\usepackage{natbib}
\usepackage{units}
\usepackage{url}
\usepackage{lscape}
\usepackage{supertabular,booktabs}

\newcommand{\hii}{\mbox{\ion{H}{\small II}}}
\newcommand{\rahms}[4]{$#1^{\rm h}#2^{\rm m}#3\mbox{$^{\rm s}\mskip-7.6mu.\,$}#4$} 
\newcommand{\decdms}[4]{$#1^{\circ}#2'#3\mbox{$''\mskip-7.6mu.\,$}#4$} 

\begin{document}

\title{The richness of compact radio sources in NGC 6334D to F}
\author{S.-N. X.\ Medina\inst{1},
S. A.\ Dzib\inst{1},
M. Tapia\inst{2},
L. F. Rodr\'{\i}guez\inst{3},
and
L. Loinard\inst{3,} \inst{4}
}

\institute{Max-Planck-Institut f\"ur Radioastronomie, Auf dem H\"ugel 69,
 D-53121 Bonn, Germany 
 \and Instituto de Astronom\'{\i}a, Universidad Nacional Aut\'onoma de M\'exico, Ensenada, B. C., CP 22830, Mexico
 \and Instituto de Radioastronom\'{\i}a y Astrof\'{\i}sica, Universidad Nacional Aut\'onoma de M\'exico, Morelia 58089, Mexico
 \and Instituto de  Astronom\'{\i}a, Universidad Nacional Aut\'onoma de M\'exico, Apartado Postal 70-264, CdMx C.P. 04510, Mexico 
 \\
\email{smedina, sdzib, @mpifr-bonn.mpg.de; mt@astrosen.unam.mx;  l.rodriguez, and l.loinard, @crya.unam.mx
 }
}

\date{Received 2017; }
\abstract
{The presence and properties of compact radio sources embedded in massive star-forming regions can reveal important physical properties about these regions and the processes occurring within them. The NGC 6334 complex, a massive star-forming region, has been studied extensively. Nevertheless, none of these studies has focused in its content in compact radio sources.}
{Our goal here is to report on a systematic census of the compact radio sources toward { NGC~6334}, and their characteristics. This will be used to try and define their very nature.}
{We use VLA C band (4--8 GHz) archive data with { 0\rlap{$''$}\,.36 (500 AU)}  of spatial resolution and noise level of 50~$\mu$Jy bm$^{-1}$ to carry out a systematic search for compact radio sources within NGC 6334. We also search for infrared counterparts to provide some constraints on the nature of the detected radio sources.}
{A total of 83 compact sources and three slightly resolved sources were detected. Most of them are here
reported for the first time. We found that 29 of these 86 sources have infrared counterparts and 
three are highly variable.  
Region~D contains 18 of these sources. The compact source toward the center, in projection, of region E is also detected.}
{From statistical analyses, we suggest that the 83 reported compact sources are real and most of them are related to NGC~6334 itself. A stellar nature for 27 of them is confirmed by their IR emission. 
Compared with Orion, region D suffers a deficit of compact radio sources.
The infrared nebulosities around two of the slightly resolved sources are suggested to be warm dust, and we argue that the associated radio sources trace free-free emission from ionized material. 
We confirm the thermal radio emission of
the compact source in region E. However, its detection at infrared wavelengths implies 
that it is located in the foreground of the molecular cloud. Finally, three strongly variable 
sources are suggested to be magnetically active young stars.} 
\titlerunning{Compact sources in NGC 6334D to F}

\maketitle

\section{Introduction}
Massive star-forming regions contain high- and low-mass stars,
which often present radio emission from different
origins. The high-mass stars may ionize
the medium around them, creating an \hii\ region, which produces free-free 
emission detectable at radio wavelengths. Depending on the size and density of 
the ionized region they can be classified as classic, compact (C), 
ultra-compact (UC), and hyper-compact (HC) \hii\ regions (Kurtz 2005). 
Also, massive stars may produce radio emission from
their ionized winds (Contreras et al. 1996) and, in massive multiple 
stellar systems, from their wind collision regions (e.g., 
Ortiz-Le\'on et al. 2011). On the other hand, young low-mass stars may 
produce free-free radio emission in their jets (Anglada 1996),
from their externally ionized disk by the UV photons of an OB star 
(proplyds; O'dell et al. 1993), and gyrosynchrotron radio emission
when they are magnetically active. Even for nearby star forming regions 
(d~$<2$~kpc), the radio emission of HC~\hii ~regions and from low-mass 
stars occurs on subarcsecond scales (see Rodr\'{\i}guez 
et~al.~2012 for a detailed  description on the characteristics of 
the different types of compact radio emission). Massive star forming 
regions contain many unresolved, or slightly resolved, radio 
sources (e.g. Orion and M17; Forbrich et al. 2016, Rodr\'{\i}guez et~al.~2012). 
The compact radio sources have long been known to exist also within extended
HII regions such as Orion, and have recently been found to be abundant in 
other cases. They are important because they might play a role in the time 
evolution of the HII regions themselves. In this work we present the first
radio detection of a large number of radio sources in NGC~6334.

The giant molecular cloud NGC~6334 is a complex with very active 
spots of massive star formation at different evolutionary stages (see Persi 
\& Tapia 2008 for a detailed review). Located at the relative 
nearby distance of 1.34$^{+0.15}_{-0.12}$ kpc (Reid~et~al.~2014)
it is an ideal target for the study of the different phases
of massive star formation and the detailed analysis of its content in  compact radio
sources may provide strong clues on the phases to which they belong. 

Early radio maps with high resolution ($\sim1''$) of NGC~6334
showed the existence of six strong radio sources which were named
as sources NGC~6334A to F (Rodr\'{\i}guez et al. 1982, from now on
we will refer to them only with the alphabetic name). 
Source B  is extragalactic (Moran et al. 1990; 
Bassani et al. 2005), while the others are compact 
and UC-\hii\ regions. In this manuscript we will
focus in the study of compact sources  around the regions D, E and F 
and their surroundings, including the region called NGC~6334I(N) 
(hereafter I(N)). These sources are located in the north-east portion of the cloud. 
\begin{figure*}[!ht]
   \centering
  \includegraphics[height=0.80\textwidth,trim= 50 30 15 15, clip]{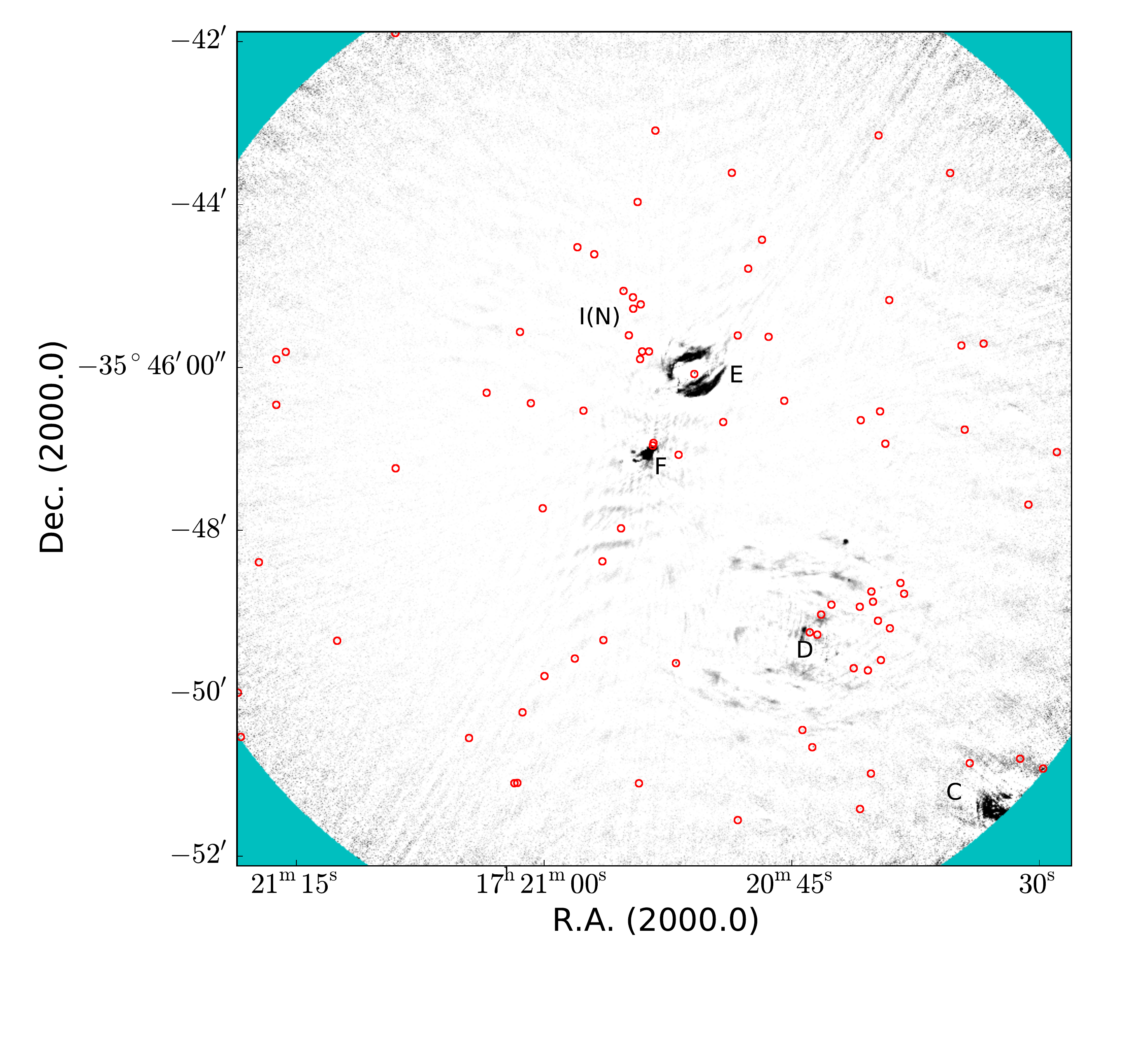}
   \caption{VLA image of NGC~6334E to F and I(N). Small circles indicate the
   position of the detected
   compact sources. }
   \label{fig:full}
\end{figure*}

 Briefly described, source D is an evolved \hii\ region with a 
 nearly circular shape of approximate radius of 75$''$ (0.5 pc),
centered on what appears to be one of  its main ionizing early B type stars, 2MASS J17204800-3549191, reddened by about  A$_V$  = 10  (Straw et al. 1989). At its western 
edge, the expansion of the \hii\ region seems to be 
halted by a dense dark cloud, { where CXOU 172031.76-355111.4 (also
known as 2MASSJ17204466-3549168 or NGC6334II-23), a
more luminous late O-type ionizing star, lies (Feigelson et al. 2009; Straw  \& Hyland 1989)}. This
interaction appears to have triggered a second star formation stage 
in the region inside the dense dark cloud
(Persi \& Tapia 2008). Regions E and F, widely studied at several 
wavelengths, are \hii\ 
regions that contain many signs of star formation (Persi \& Tapia 2008 
and references therein).
{The C-\hii\ region E} shows an extended shell-like structure with a compact radio source 
at its center (Carral et al. 2002).
Source F is a younger and complex UC~\hii\ region with a cometary shape 
and a radio flux of 
$\sim~3$~Jy. This source and its surroundings are collectively known as 
NGC~6334I. Finally, the region
I(N) is the youngest of all, as it has just started to form massive 
stars (Rodr\'{\i}guez et al. 2007;
Hunter et al. 2006).

\section{Observations}

\begin{figure*}[!th]
   \centering
  \includegraphics[height=0.75\textwidth,trim= 20 20 40 50, clip]{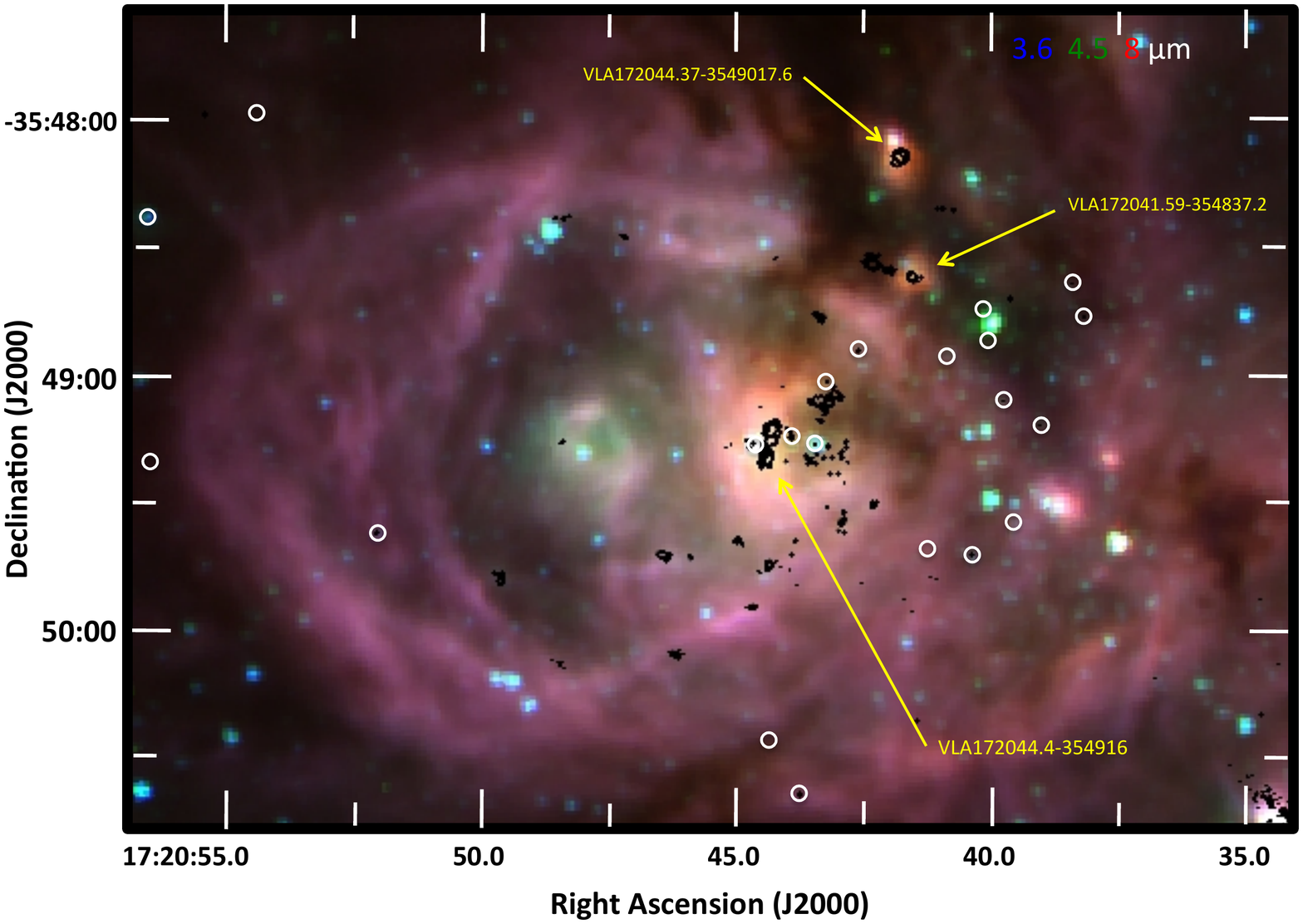}
   \caption{Composite SPITZER/GLIMPSE infrared image of NGC~6334D. Small circles indicate the 
   position of the detected compact radio sources. Contours trace the radio 
   continuum image at 6.0 GHz, and the contour levels are at 90, 180, 300, 450,
   750, and 1500~$\mu$Jy/beam. Slightly extended radio sources are indicated.}
   \label{fig:D}
\end{figure*}

We will use an archival observation of NGC~6334 obtained with the 
Karl G. Jansky Very Large Array (VLA)
telescope of the NRAO\footnote{The National Radio Astronomy Observatory 
is operated by Associated Universities Inc. under cooperative agreement 
with the National Science Foundation.} in the C-band (4 to 8 GHz) 
obtained as part of the project 10C-186. The data were taken on 7 July 
2011, while the array was in A-configuration, its most extended. Two 
sub-bands, each 1 GHz wide and centered at 5.0 and 7.1 GHz, respectively, 
were recorded simultaneously. The average frequency of the whole observation
is 6.0 GHz. The quasars J1717$-$3342, J1924$-$2914 and J1331+3030, were 
used as the gain, bandpass, and flux calibrators, respectively. The total 
time spent on source was 84 minutes. First results of this observation 
were reported by Hunter et al. (2014) and Brogan et al. (2016), we point 
the reader to these papers for further details of the observation.

The data were edited, calibrated and imaged using the software CASA with the 
help of the VLA Calibration Pipeline. The image at 6.0 GHz was produced by combining the two
sub-bands using a multi-frequency deconvolution software (e.g., Rau \& Cornwell
2011) and with a pixel size of 0\rlap{.}$''$06. The resulting beamsize is
0\rlap{.}$''67\times$0\rlap{.}$''$19; P.A. = $-2\rlap{.}^\circ$7.
The used weighting scheme was intermediate between natural and uniform (robust = 0, Briggs 1995).
Similar images were also produced for both sub-bands separately, with pixel 
sizes of 0\rlap{.}$''$06 and 0\rlap{.}$''$04 for 5.0 GHz and 7.1 GHz, respectively. The beamsize for the 5.0 GHz image is  
0\rlap{.}$''91\times$0\rlap{.}$''$25; P.A. = $-2\rlap{.}^\circ$3,
and
0\rlap{.}$''66\times$0\rlap{.}$''$17; P.A. = $-2\rlap{.}^\circ$8
for the image at 7.1 GHz. { All the images are corrected for the 
primary beam response.}

\section{Source extraction}

The source extraction was performed using the BLOBCAT software (Hales et al. 2012). 
This software is a flood fill algorithm that  cataloges islands of agglomerated 
pixels (blobs) within locally varying noise. It is designed for two-dimensional 
input FITS images of surface brightness (SB). BLOBCAT makes improvements in the 
morphological assumptions and  applies bias corrections to extract blob properties 
(see Hales et al. 2012 for details). To run BLOBCAT, we also use the rms estimator 
algorithm implemented within the SExtractor package (Bertin \& Arnouts 1996, Holwerda 2005) 
to make a suitable noise map of the SB image. We use a mesh size of 60$\times$60 pixels 
and  Signal to Noise Ratio (SNR) $=$  5 as threshold for detecting blobs for the 6.0 GHz 
image. For the 5.0 and 7.1 GHz images the mesh size was 80$\times$80 pixels and SNR = 3 
as threshold, but in this case only sources that were also founded in the 6.0 GHz
map were considered as real. The previous calculations were made following the derivations 
of Hales et al. (2012).

The expected number of false detections is calculated using the complementary cumulative 
distribution function  $\Phi(x)=1- \phi(x)$ where $\phi(x)$ is the  cumulative distribution 
function. Assuming that the noise in our radio maps 
follows a Gaussian distribution
\begin{equation} 
\phi(x)=\frac{1}{2} \bigg [  1 + erf \bigg( \frac{x}{\sqrt{2}}  \bigg) \bigg ]
\end{equation}
where erf is the error function given by 
\begin{equation} 
erf(x)= \frac{1}{\sqrt{\pi}} \int_{-x}^{x} e^{-t^2} dt.
\end{equation}
$\Phi(x)$ is the probability that a value of a standard normal random variable X will exceed 
an $x$ level. So, the probability that any independent pixel (synthesized beam) will have a 
value up to 5$\sigma$ is $\Phi(5)  \approx 3\times10^{-7}$. In consequence, 
we expect a total of 1 source above 5$\sigma$ in our 6 GHz maps, and conclude that essentially 
all our sources catalogued in the region are real.
In consequence, the sources above 5$\sigma$ at 6.0 GHz are trustworthy. { Even so, we
did a visual inspection of all these sources to confirm their detection. We excluded the artifacts located close to the edge of the image.}


\section{Results}
\subsection{Radio sources}
We obtain an image of $10'\times10'$, which is displayed in Figure~\ref{fig:full},  
combining the two observed sub-bands. The noise level is 
position dependent and produces the following two effects. The first is that  
the noise increases at the edges of the image, which is expected from the Gaussian 
primary beam pattern. The second is due to the imperfect
sampling of the UV-space and it mainly affects areas around the extended strong emission. 
Thus, the noise level is close to 50~$\mu$Jy near the \hii\ regions, but only 
about 8 $\mu$Jy in areas free of extended emission and not far from the center 
of the image. A similar effect is present in the Orion observations of Forbrich 
et al. (2016).

\longtab{
\setlength{\tabcolsep}{8pt}
\renewcommand{\arraystretch}{1.51}
\begin{longtable}{p{0.3cm}p{3.4cm}p{4.1cm}p{1.4cm}p{1.4cm}p{1.5cm}p{1.7cm}p{0.9cm}}
  \caption{\label{tab:RS} Radio properties of compact sources in NGC~6334}\\
\hline\hline
ID &          & Other  & $S_\nu(6\,{\rm GHz})$ & $S_\nu(5\,{\rm GHz})$ &$S_\nu(7\,{\rm GHz})$ &    &      \\
\# & VLA Name   & Name   &($\mu$Jy) & ($\mu$Jy) &($\mu$Jy) & $\alpha$ & SNR\\
\hline
\endfirsthead
\caption{continued.}\\
\hline\hline
ID & & Other
  & $S_\nu(6\,{\rm GHz})$ & $S_\nu(5\,{\rm GHz})$ &$S_\nu(7\,{\rm GHz})$ &    &      \\
\# &VLA Name   & Name   &($\mu$Jy) & ($\mu$Jy) &($\mu$Jy) & $\alpha$ & SNR\\
\hline
\endhead
\hline
\hspace{0.6pt}
\endfoot

\multicolumn{8}{l}{Sources inside region D}\\

\hline
1&J172038.21-354846.8&...&	372$\pm$24&	421$\pm$27&	305$\pm$26&	-0.9$\pm$0.3&	28.7\\
2&J172038.43-354838.9&...&	118$\pm$14&	141$\pm$17&	81$\pm$19&	-1.6$\pm$0.7&	9.4\\
3&J172039.07-354912.3&...&	127$\pm$15&	125$\pm$17&	168$\pm$23&	0.9$\pm$0.6&	9.6\\
4&J172039.61-354935.7&...&	80$\pm$16 &	89$\pm$18&	$<$66$\pm$22&	$<-$0.9$\pm$1.1&5.2\\
5&J172039.79-354906.7&...&	144$\pm$17&	176$\pm$20&	116$\pm$21&	-1.2$\pm$0.6&	9.6\\
6&J172040.09-354852.7&...&	63$\pm$13&70$\pm$17&	$<$63$\pm$21 &	$<-$0.3$\pm$1.2&	5.1\\
7&J172040.19-354845.2&...&	76$\pm$15&	85$\pm$19&	98$\pm$11&	0.4$\pm$0.7&	5.7\\
8&J172040.40-354943.3&2MASS J17204040-3549438&	168$\pm$17&	143$\pm$18&	223$\pm$26&	1.3$\pm$0.5&	11.8\\
9&J172040.89-354856.4&...&	81$\pm$13&	84$\pm$17&	75$\pm$17&	-0.3$\pm$0.9&	6.3\\
10&J172041.26-354941.7&...&	77$\pm$15&	95$\pm$19&	$<$66$\pm$22&	$<-$1.0$\pm$1.1&	5.5\\
11&J172042.61-354854.9&2MASS J17204261-3548550&	210$\pm$24&	185$\pm$28&	239$\pm$22&	0.7$\pm$0.5&	9.7\\
12&J172043.23-354902.2&...&	236$\pm$28&	269$\pm$30&	189$\pm$20&	-1.0$\pm$0.4&	9.6\\
13&J172043.46-354917.0&2MASS J17204345-3549172&	271$\pm$33&	266$\pm$44&	212$\pm$23&	-0.6$\pm$0.6&	9.3\\
14&J172043.92-354915.3&...&	568$\pm$43&	612$\pm$50&	522$\pm$34&	-0.5$\pm$0.3&	18.4\\
15&J172044.66-354916.7&2MASS J17204466-3549168&	129$\pm$12&	120$\pm$13&	126$\pm$13&	0.1$\pm$0.4&	5.1\\

\hline\hline
\multicolumn{8}{l}{Remaining Sources}\\
\hline
16&J172028.97-354702.4&...&	109$\pm$22&	$<$105$\pm$35&	...&	...&	5.1\\
17&J172029.78-355055.5&...&	305$\pm$60&	210$\pm$42&	...&	...&	5.7\\
18&J172030.69-354741.1&...&	95$\pm$19&	73$\pm$19&	...&	...&	5.6\\
19&J172031.18-355048.3&...&	232$\pm$48&	116$\pm$34&	...&	...&	5.0\\
20&J172033.41-354542.5&...&	85$\pm$17&	72$\pm$18&	...&	...&	5.5\\
21&J172034.23-355051.6&...&	793$\pm$57&	772$\pm$52&	...&	...&	20.8\\
22&J172034.55-354645.9&...&	70$\pm$14&	64$\pm$17&	...&	...&	5.0\\
23&J172034.75-354543.9&...&	70$\pm$14&	54$\pm$17&	...&	...&	5.3\\
24&J172035.44-354336.9&...&	114$\pm$23&	91$\pm$22&	...&	...&	5.1\\
25&J172039.12-354510.5&2MASS J17203909-3545108&	63$\pm$12&	71$\pm$15&	75$\pm$20&	0.2$\pm$1.0&	5.7\\
26&J172039.35-354656.3&...&	49$\pm$10&	66$\pm$15&	$<$51$\pm$17&	$<-$0.7$\pm$1.2&	5.0\\
27&J172039.67-354632.5&...&	90$\pm$10&	98$\pm$13&	77$\pm$15&	-0.7$\pm$0.7&	9.7\\
28&J172039.77-354309.2&...&	106$\pm$21&	80$\pm$22&	...&	...&	5.3\\
29&J172040.21-355059.3&...&	131$\pm$21&	103$\pm$21&	...&	...&	6.5\\
30&J172040.84-354639.0&...&	68$\pm$10&	84$\pm$13&	53$\pm$13&	-1.3$\pm$0.8&	7.4\\
31&J172040.87-355125.4&...&	130$\pm$26&	106$\pm$24&	...&	...&	5.1\\
32&J172043.76-355039.9&...&	877$\pm$50&	976$\pm$56&	...&	...&	53.8\\
33&J172044.36-355027.2&...&	76$\pm$15&	77$\pm$19&	...&	...&	5.2\\
34&J172045.47-354624.7&...&	41$\pm$08&	38$\pm$11&	41$\pm$11&	0.2$\pm$1.1&	5.3\\
35&J172046.42-354537.6&...&	83$\pm$09&	73$\pm$12&	95$\pm$12&	0.8$\pm$0.6&	10.1\\
36&J172046.82-354426.1&...&	53$\pm$10&	$<$60$\pm$20&	$<$48$\pm$16&	...&	5.2\\
37&J172047.65-354447.4&...&	52$\pm$10&	41$\pm$13&	59$\pm$14&	1.0$\pm$1.1&	5.3\\
38&J172048.27-355133.6&...&	126$\pm$21&	109$\pm$21&	...&	...&	6.7\\
39&J172048.28-354536.6&SSTU J172048.27-354537.0&	80$\pm$11&	98$\pm$17&	55$\pm$13&	-1.7$\pm$0.8&	7.8\\
40&J172048.64-354336.8&...&	62$\pm$13&	71$\pm$15&	$<$69$\pm$23&	$<-$0.1$\pm$1.2&	5.0\\
41&J172049.16-354640.4&...&	41$\pm$08&	40$\pm$12&	40$\pm$12&	0.0$\pm$1.2&	5.4\\
42&J172050.91-354605.0&2MASS J17205087-3546047&	821$\pm$51&	612$\pm$64&	889$\pm$52&	1.1$\pm$0.3&	32.5\\
43&J172051.86-354704.5&CM1&	85$\pm$12&	72$\pm$13&	104$\pm$16&	1.1$\pm$0.7&	7.4\\
44&J172052.02-354938.0&CXOU 172052.0-354938&	878$\pm$48&	834$\pm$47&	978$\pm$55&	0.5$\pm$0.2&	87.4\\
45&J172053.26-354305.7&[S2000e] SM6&	90$\pm$15&	105$\pm$17&	...&	...&	6.4\\
46&J172053.38-354655.8&CM2&	256$\pm$41&	266$\pm$75&	233$\pm$21&	-0.4$\pm$0.9&	6.6\\
47&J172053.42-354657.7&[HBM2006b] I-SMA1&	466$\pm$70&	551$\pm$107&	608$\pm$131&	0.3$\pm$0.8&	7.1\\
48&J172053.65-354548.4&...&	258$\pm$16&	230$\pm$17&	307$\pm$19&	0.8$\pm$0.3&	31.2\\
49&J172054.06-354548.4&SSTU J172053.96-354547.6&	54$\pm$09&	50$\pm$10&	54$\pm$10&	0.2$\pm$0.8&	6.5\\
50&J172054.15-354513.7&[HBC2014] H$_2$O-C4&	98$\pm$10&	148$\pm$13&	75$\pm$11&	-1.1$\pm$0.4&	11.4\\
51&J172054.19-354554.0&...&	61$\pm$09&	70$\pm$12&	73$\pm$10&	0.1$\pm$0.6&	7.5\\
52&J172054.26-355106.5&...&	80$\pm$16&	63$\pm$18&	...&	...&	5.3\\
53&J172054.34-354358.3&SSTU J172054.33-354358.4&	74$\pm$10&	71$\pm$14&	84$\pm$17&	0.5$\pm$0.8&	7.6\\
54&J172054.60-354516.9&...&	45$\pm$08&	51$\pm$12&	60$\pm$11&	0.4$\pm$0.8&	5.7\\
55&J172054.62-354508.5&[HBM2006b] I(N)-SMA4&	162$\pm$12&	152$\pm$14&	181$\pm$14&	0.5$\pm$0.3&	19.1\\
56&J172054.87-354536.5&[HBC2014] VLA 2&	86$\pm$10&	98$\pm$12&	64$\pm$11&	-1.2$\pm$0.6&	10.1\\
57&J172055.19-354503.8&[HBM2006b] I(N)-SMA1b&	353$\pm$21&	347$\pm$22&	360$\pm$22&	0.1$\pm$0.3&	43.7\\
58&J172055.34-354758.7&...&	226$\pm$17&	212$\pm$21&	248$\pm$18&	0.5$\pm$0.4&	18.7\\
59&J172056.41-354921.0&...&	50$\pm$10&	$<$60$\pm$20&	$<$48$\pm$16&	...&	5.1\\
60&J172056.47-354823.0&CXOU J172056.4-354823&	55$\pm$10&	58$\pm$12&	62$\pm$11&	0.2$\pm$0.8&	5.5\\
61&J172056.96-354436.8&...&	46$\pm$09&	43$\pm$12&	66$\pm$13&	1.2$\pm$1.0&	5.2\\
62&J172057.62-354632.0&...&	76$\pm$08&	105$\pm$13&	52$\pm$10&	-2.0$\pm$0.7&	10.4\\
63&J172057.98-354431.6&...&	252$\pm$16&	272$\pm$19&	236$\pm$19&	-0.4$\pm$0.3&	27.6\\
64&J172058.14-354934.6&...&	625$\pm$35&	689$\pm$39&	560$\pm$33&	-0.6$\pm$0.2&	62.9\\
65&J172059.98-354947.5&...&	55$\pm$10&	60$\pm$10&	$<51\pm17$&	$<-$0.5$\pm$1.1&	5.6\\
66&J172100.08-354743.9&...&	41$\pm$08&	45$\pm$11&	43$\pm$10&	-0.1$\pm$1.0&	5.4\\
67&J172100.80-354626.5&...&	157$\pm$11&	157$\pm$13&	153$\pm$13&	-0.1$\pm$0.3&	21.2\\
68&J172101.31-355014.2&...&	64$\pm$13&	65$\pm$14&	$<$60$\pm$20&	$<-$0.2$\pm$1.1&	5.1\\
69&J172101.45-354534.0&...&	41$\pm$08&	58$\pm$11&	$<$60$\pm$20&	$<-$1.9$\pm$1.1&	5.1\\
70&J172101.62-355106.2&...&	102$\pm$19&	96$\pm$20&	...&	...&	5.5\\
71&J172101.80-355106.4&...&	232$\pm$22&	221$\pm$22&	...&	...&	12.4\\
72&J172103.47-354618.8&...&	573$\pm$32&	639$\pm$36&	506$\pm$29&	-0.7$\pm$0.2&	73.8\\
73&J172104.10-354540.5&...&	44$\pm$09&	47$\pm$12&	38$\pm$12&	-0.6$\pm$1.2&	5.0\\
74&J172104.55-355033.1&...&	94$\pm$17&	70$\pm$12&	...&	...&	5.7\\
75&J172108.98-354714.4&...&	76$\pm$11&	57$\pm$13&	93$\pm$16&	1.4$\pm$0.8&	7.4\\
76&J172108.98-354153.8&...&	309$\pm$62&	208$\pm$43&	...&	...&	5.1\\
77&J172112.53-354921.4&...&	92$\pm$18&	83$\pm$22&	...&	...&	5.1\\
78&J172115.62-354548.6&...&	94$\pm$18&	70$\pm$18&	...&	...&	5.3\\
79&J172116.19-354554.1&...&	98$\pm$19&	75$\pm$18&	...&	...&	5.3\\
80&J172116.20-354627.6&...&	97$\pm$19&	78$\pm$19&	...&	...&	5.3\\
81&J172117.26-354823.5&...&	120$\pm$24&	85$\pm$22&	...&	...&	5.1\\
82&J172118.37-355032.2&...&	321$\pm$62&	207$\pm$43&	...&	...&	5.3\\
83&J172118.54-354959.6&...&	274$\pm$55&	$<$183$\pm$61&	...&	...&	5.2\\
\hline\noalign{\smallskip}


\multicolumn{8}{l}{\tablefoot{Other names are obtained from the following catalogs: 2MASS = Cutri et al. (2003); CXOU = Feigelson et al. (2009); 
[HBM2006b] = Hunter et al. (2006); CM = Brogan et al. (2016); SSTU = Willis et al. (2013); [S2000e] = Sandell (2000); and  [HBC2014] = Hunter et al. (2014).  The SNRs are obtained from the 6.0~GHz map. 
Upper-limits to fluxes are obtained as three times the noise level in the image, that are at the same time used as the error for these  upper-limits.}}
\end{longtable}

}

In Figure~\ref{fig:full}, the UC~\hii\ region F and  C~\hii\ region E are immediately appreciated, while only
a fraction of the extended emission of source D is recovered. Only a small
portion of region C falls inside the primary beam of this observation.
Additional to the 
extended emission of the \hii\ regions, a total 
of 83 compact\footnote{Compact sources are defined here as source whose deconvolved 
sizes are similar to, or smaller than 
the synthesized beam.} radio sources were also detected (Table \ref{tab:RS}), including 
those previously reported in the E, I, and I(N) region. The compact sources cannot be
easily appreciated in this large field figure and they are represented by small red
circles. The astrometric accuracy for the positions is better than 0\rlap{$''$}\,.1,
and we have used the positions of the sources to name them according to the IAU
suggestion. 

Using equation A11 from Anglada et al. (1998) the expected number of background 
extragalactic sources in the imaged area with flux densities above 50 $\mu$Jy is 
7$\pm3$. This previously computed number only reflects an upper limit because the emission 
of the extended sources affects the noise distribution and this changes the value of 
the expected background sources. Most of the detected compact radio sources 
are Galactic objects and most probably related to the NGC 6334 complex.

Region D, presented in Figure \ref{fig:D}, contains a significant fraction of 
all of the detected compact radio sources. Also, at the
north side of this region there are two cometary shaped radio sources and at its center 
a double source. All of them, the compact
radio sources, the cometary radio sources and the double radio sources, are reported 
here for the first time.

In order to compute the spectral index ($\alpha$; $S_{\nu}\propto\nu^{\alpha}$), 
we measured the flux density of the detected sources in the 5.0 GHz and 7.1 GHz
maps and calculated the slope between both points. The error was calculated by 
using the standard error propagation theory. As the field of view of the map
at 7.1~GHz is smaller than at 5.0~GHz, we could not calculate the spectral index
for some sources that lie outside the edges of the 7.1 GHz image. 
Additionally, some sources were detected in one band but not in 
the other, in those cases we just calculate a limit for $\alpha$.
The spectral index for each individual source is shown in Table~\ref{tab:RS}.

{ Using the SIMBAD database we searched for counterparts of detected radio sources.
Inside a radius of 0\rlap{.}$''$5 from the VLA position, we found 17 radio sources with 
counterparts at different wavelengths and are shown in Table~\ref{tab:RS}. Additionally, 
the source SSTU J172053.96-3545.6 is at an angular 
distance of 0\rlap{.}$''$9 from VLA J172054.06-354548.4. This shift is larger than  
the combined position errors of both telescopes (pointing accuracy for IRAC is $\sim$0\rlap{.}$''$5)
and indicates that these sources are not the same object, but we cannot discard that they 
may be related (e.g., a binary system). Finally, source [S2000e]~SM6 is at an angular distance from
source VLA~J172053.26-354305.7 of $\sim$1\rlap{.}$''$2, this agrees within the position error  of 2$''$ for the millimeter 
source (Sandell 2000). To look for more infrared counterparts we perform a more detailed search, that is described
in the next section.}


\subsection{Infrared counterparts}
\begin{figure}[!th]
   \centering
  \includegraphics[height=0.60\textwidth,trim= 70 30 20 160, clip]{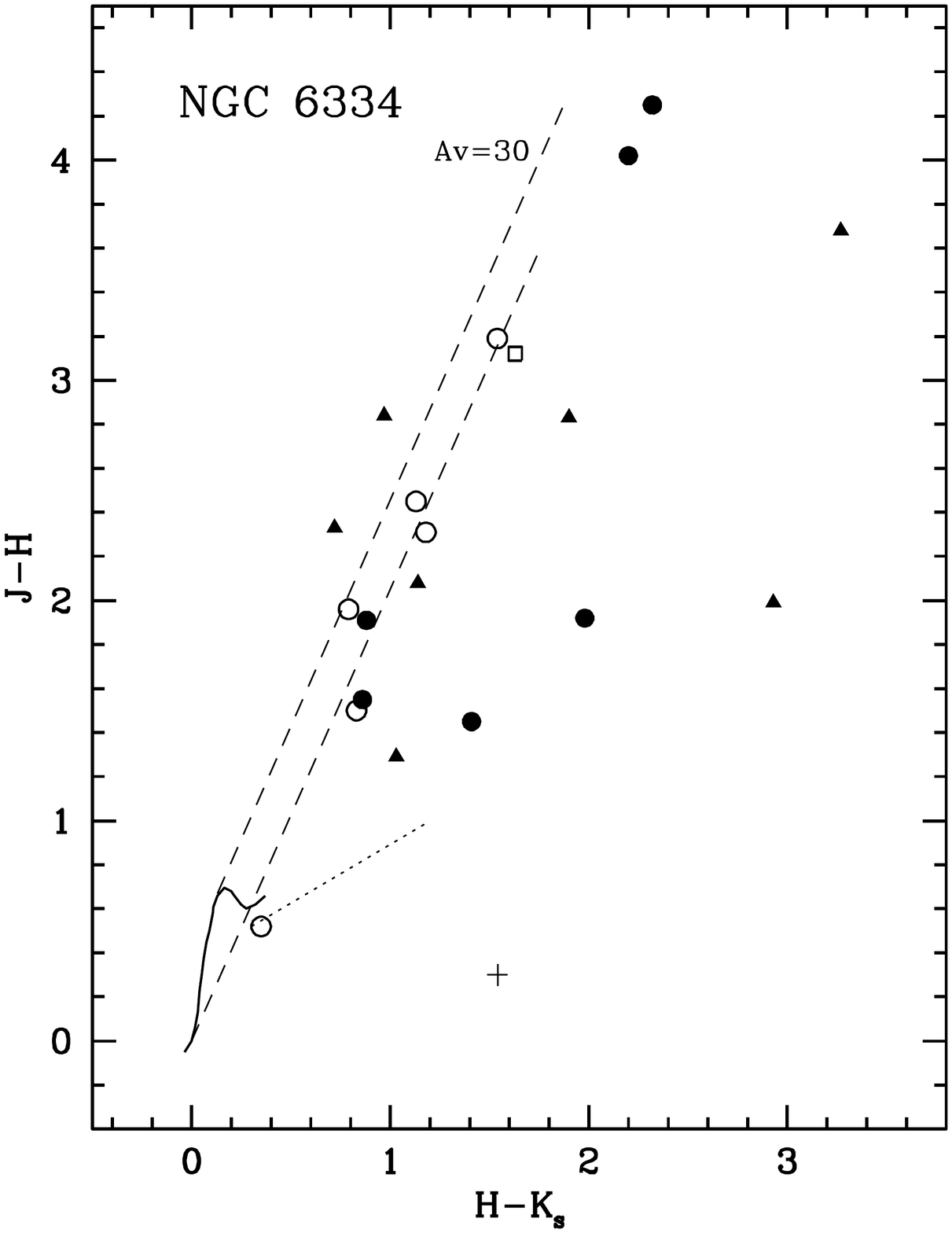}
   \caption{ $J-H$ versus $H - K_s$ diagram of the near-IR counterparts with 
   measurements in the three bands. Solid circles are sources
in the dark cloud west of the HII region D. Solid triangles correspond to sources 
within the limits of CNN. Open squares are
in region I(N). Open circles are for sources in the rest of the observed field. 
The solid line marks the locus  of 
the main sequence (Koornneef 1983), the dashed lines delineate the reddening 
band for all main sequence star and 
giant stars (Rieke and Lebofsky 1985). { The small plus sign near the bottom of the plot illustrates the maximum photometric errors.}
   }
   \label{fig:3}
\end{figure}

\begin{figure}[!bh]
   \centering
  \includegraphics[height=0.620\textwidth,trim= 20 150 30 150, clip]{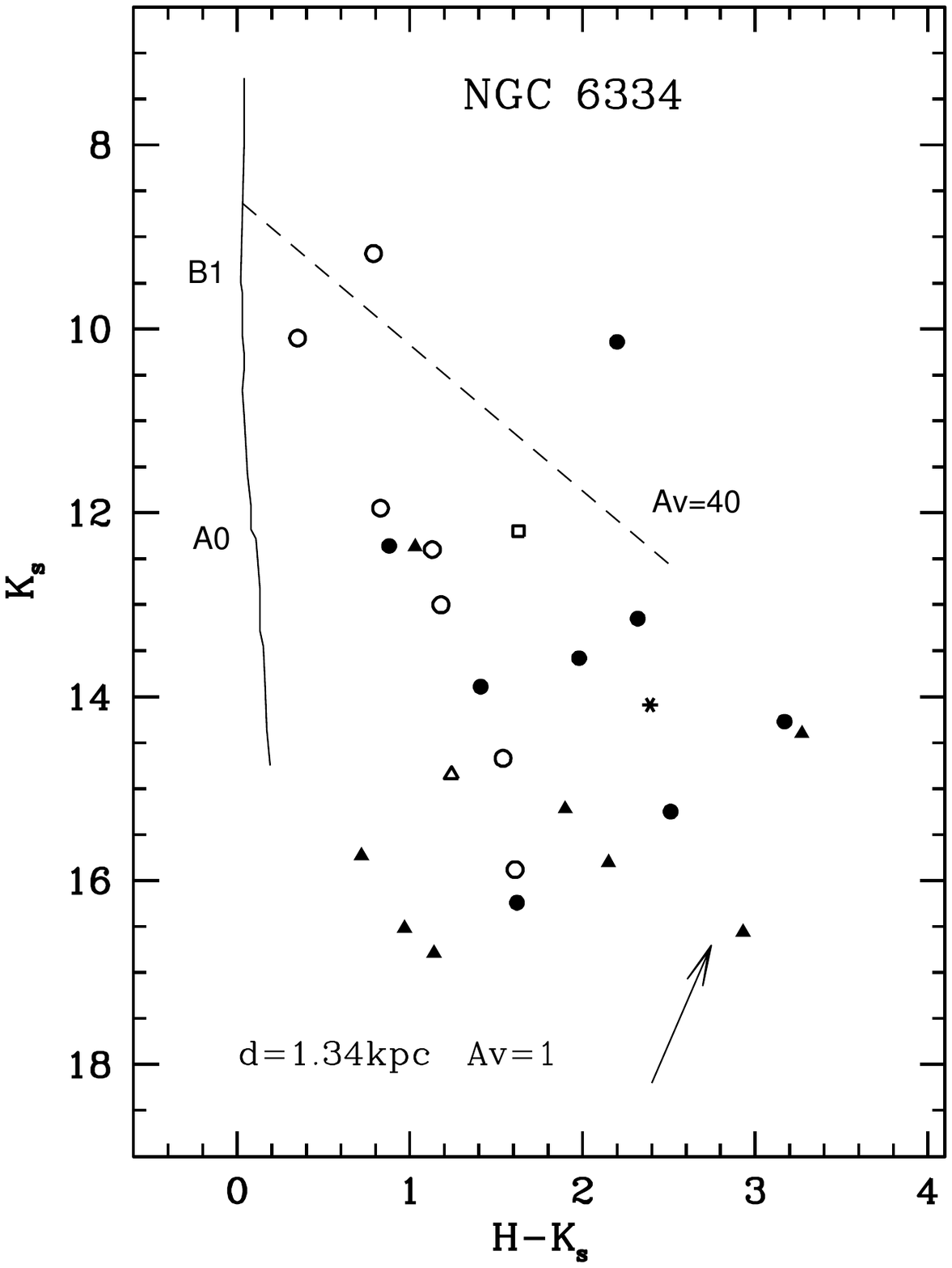}
   \caption{ $K_s$ versus $H-K_s$ diagram of the near-IR counterparts with measurements in the $H$ and $K_s$ bands. { The open triangle refers to Mir-4 in DNS and the rest of the symbols are as in Fig.~\ref{fig:3}.} For reference, the almost vertical solid line delineates the zero-age main sequence (ZAMS; Drilling \& Landolt 2000) 
for $d = 1.34$~kpc and $A_V = 1.0$. The dashed lines are the reddening vectors of length $A_V=40$ for B0 ZAMS stars.
The arrow represents the average slope of the near-IR emission excess caused by discs around YSOs, as determined by 
L\'opez-Chico \& Salas (2007). 
   }
   \label{fig:4}
\end{figure}

\begin{figure}[!th]
   \centering
  \includegraphics[height=0.450\textwidth,trim= 30 20 10 280, clip]{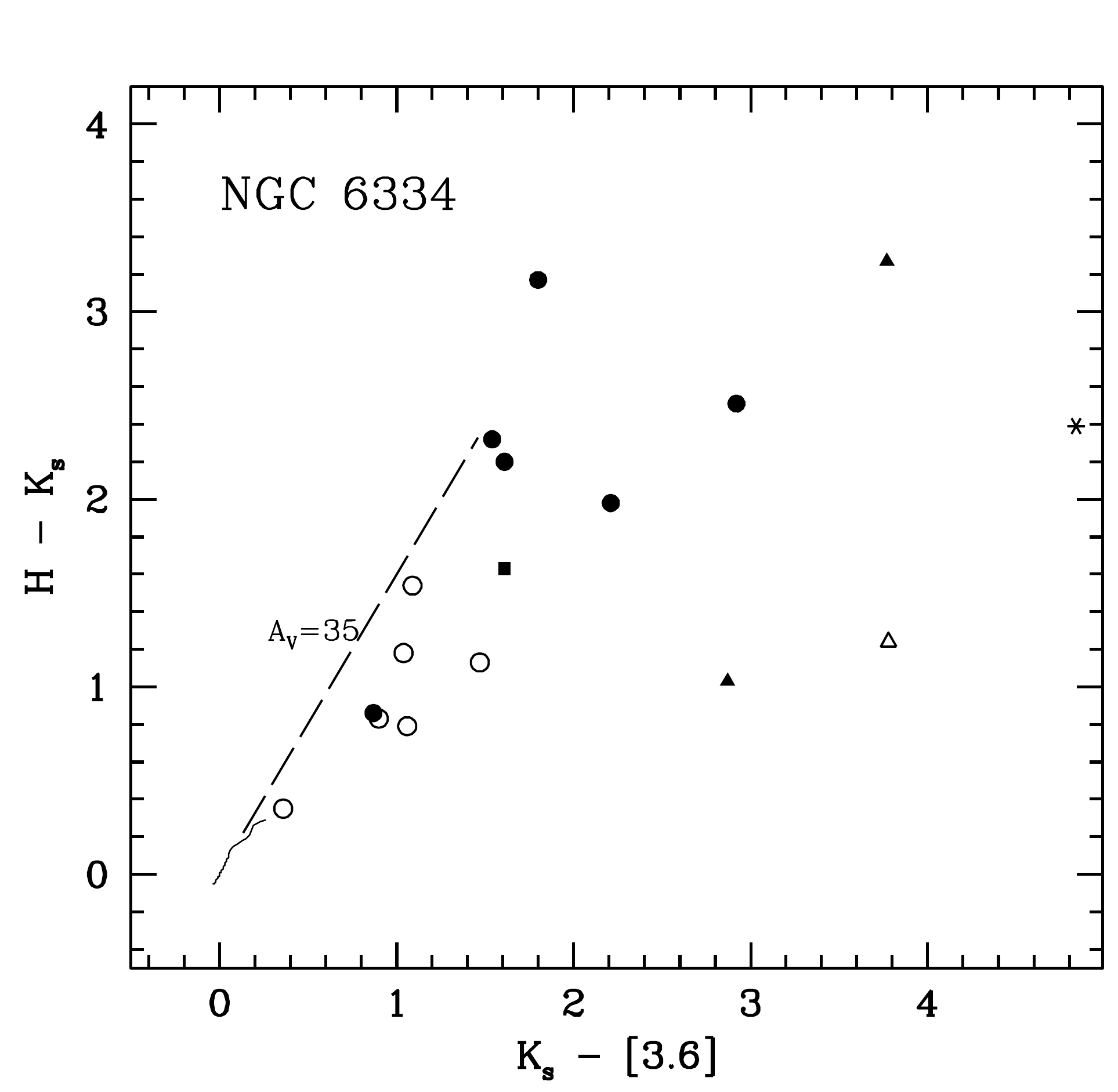}
   \caption{ $H - K_s$ versus $K_s - [3.6]$ of the IR-counterparts with measurements in the $H, K$ and IRAC channel bands.
the solid square corresponds to the central C-HII region E, { the asterisk to source \#43 in region F}. The rest of
the symbols are as in Fig.~\ref{fig:3} and~\ref{fig:4}. The solid line close to the origin represents the locus of the  main sequence 
(Koornneef 1983) and the dashed line represents the reddening vector of length $A_V=35$ (Tapia 1981).
   }
   \label{fig:5}
\end{figure}

\begin{figure}[!bh]
   \centering
  \includegraphics[height=0.470\textwidth,trim= 40 20 20 260, clip]{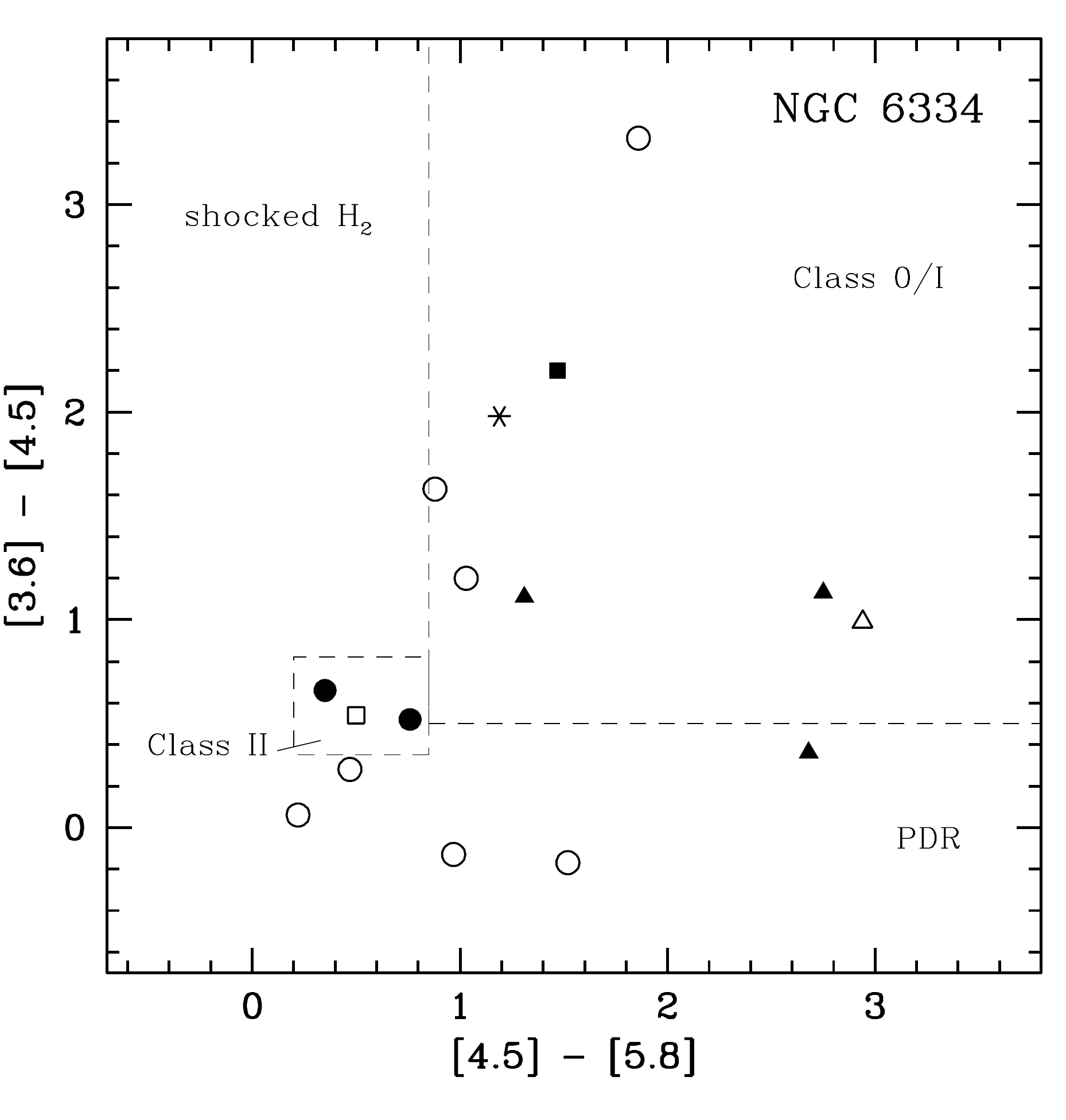}
   \caption{[3.6] - [4.5] versus [4.5] - [5.8]  diagram of the IR counterparts measured in the three IRAC bands. Symbols
are as in Figs.~\ref{fig:3} and~\ref{fig:5}. The dashed rectangles mark the locii of the Class~II and Class~0/I 
sources and the labels ``shocked H$_2$''  and ``PDR'' mark, respectively, the areas occupied by shocked 
regions emitting molecular hydrogen lines and PAH-emission-dominated photodissociation regions 
(Ybarra et al. 2014).
   }
   \label{fig:6}
\end{figure}

We searched for infrared counterparts of the compact radio sources searched
in section 4.1. We use  the source catalog from the {\it Spitzer} space 
mission as well as published near- and mid-IR photometry data by Willis 
et al. (2013) and Tapia, Persi \& Roth (1996). { We complemented the search with unpublished $JHK_s$ photometric data from M. Tapia (in preparation). 
Defining reliable criteria for assigning a positive infrared counterpart
to a compact radio source is not straightforward, as we are dealing with 
several infrared sets of infrared photometric measurements, namely from 
{\it Spitzer} images in the mid-IR and from ground-based observations with three different telescopes (and set-ups) in the near-IR. For the present 
work, we cross-checked the coordinates of the VLA compact sources with those 
from the {\it Spitzer} and also from the ground-based images. We selected those with coordinates coinciding (in all wavelength ranges) within 0\rlap{.}$''$9 or less 
to obtain a list of candidate counterparts. We then examined by eye each 
source on all available individual images (i.e., each wavelength) and checked 
for the consistency of the corresponding flux measurements. 
The mean differences in the radio, mid- and near-IR source positions for 
the 27 counterparts were -0\rlap{.}$''$07 in right ascension and +0\rlap{.}$''$07 
in declination, with standard deviations of 0\rlap{.}$''$44 and 0\rlap{.}$''$30, 
respectively. In all, no systematic coordinate offsets were found in the 
(small) area covered by this survey.

From the 83 VLA sources in Table 1, only 27 IR counterparts were found. 
Interestingly, 10 of these infrared sources are inside the dark cloud 
in region D. The ground-based near-IR and {\it Spitzer}/IRAC mid-IR 
photometry of all these sources are listed in Table~\ref{tab:IR}. 
The diagnostic two-color and color-magnitude diagrams are presented in 
Figs.~\ref{fig:3}, ~\ref{fig:4}, ~\ref{fig:5}, and ~\ref{fig:6}. The $J-H$ versus $H-K_s$ 
and the $H-K_s$ versus $K_s$-[3.6] diagrams are accurate tools for discriminating reddened photospheres (those
lying along the reddening vectors) from stars with disks (i.e.
near-IR excesses). The latter stars would lie shifted towards redder
colors in these diagrams. The amount of dust
extinction (i.e., the value of $A_V$ along the line of sight) for 
the former set of stars, is determined by the color-excess indices, under 
a ``standard'' reddening law, represented by the reddening vectors. 
References for the intrinsic colors and reddening law assumed for determining 
these parameters are given in the figure captions. 

The color-magnitude plot $K_s$ versus $H-K_s$, on the other hand, is 
the best tool for estimating approximate spectral types and intra-cloud 
extinction values for embedded stellar sources located at the distance of 
the star formation complex ($d = 1.34$ kpc and foreground value of 
$A_V = 1.0$ in this case). Finally, the IRAC two-colour magnitude [3.6]-[4.5] versus
[4.5]-[5.8] diagram provides a simple diagnostics for classifying the
evolutionary status of YSOs and of emission-line-dominated
regions (mainly Polycyclic Aromatic Hydrocarbons, PAHs,
or shocked molecular hydrogen). This is described in detail by
Ybarra et al. (2014) and references therein. The combined results
for these analyses in terms of the derived properties of
a number of individual sources are listed in the last column
of Table 2, where we also indicate whether it was detected as an
X-ray source (within 0\rlap{.}$''$6) by Feigelson et al. (2009). }

\begin{figure}[!th]
   \centering
  \includegraphics[height=0.68\textwidth,trim= 0 0 5 0, clip]{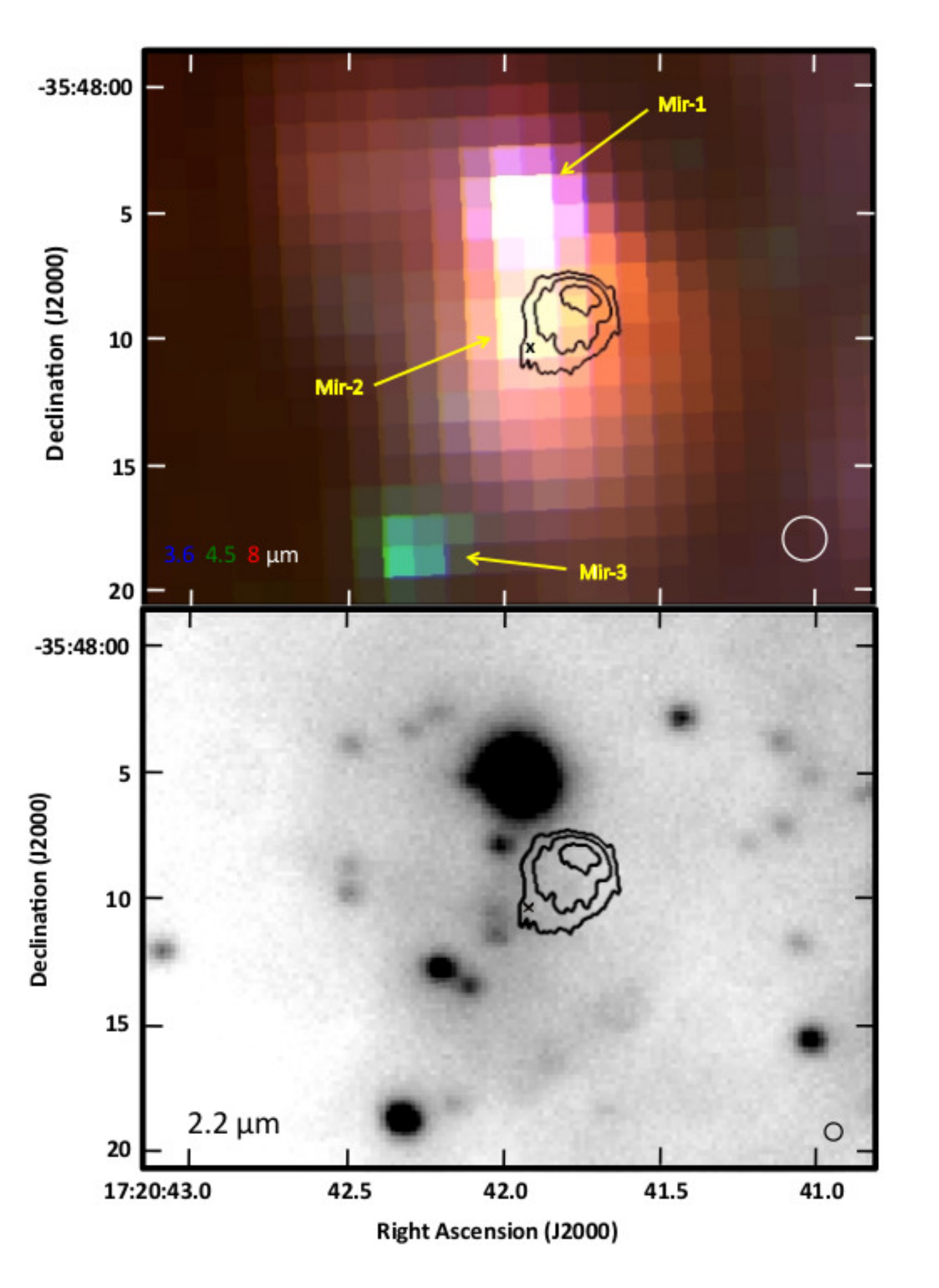}
   \caption{ Top: Composite color image of the infrared nebula coincident with VLA172041.75-3548082 (CNN) in the IRAC
bands centered at 3.6 (blue), 4.5 (green) and 8 $\mu$m (red). Superposed in black are the 6.0 GHz contours (levels are the same as Fig.~\ref{fig:D}) with the
small cross marking the position of the compact source. The three brightest mid-IR unresolved
sources are labeled (Table ~\ref{tab:IRcome}). The white circle represents the mean resolution of the IRAC images.
Bottom: Same as 7--Top, showing a higher-resolution ($0.7''$) 2.2 um image taken with the DuPont telescope
at Las Campanas Observatory (Tapia et al., in preparation).
   }
   \label{fig:CN}
\end{figure}

\begin{table*}[!th]
\footnotesize
\renewcommand{\arraystretch}{1.51}
  \begin{center}
  \caption{Infrared photometry for stellar counterparts of compact radio sources. Units of the IR values are magnitudes.}
 \begin{tabular}{ccccccccr}\hline\hline
 ID &    J   &         H   &         K   &      [3.6] &       [4.5]    &    [5.8]   &      [8]    &     Notes \\
\hline
 \multicolumn{9}{l}{Sources inside region D}\\
\hline
 3 &    --             & 17.76 $\pm$ 0.04  &  15.25 $\pm$ 0.02 &  12.33 $\pm$ 0.18 &  12.41 $\pm$ 0.19   &  --          &  --    &      2,4,9,X\\ 
 4 &  16.75 $\pm$ 0.02 & 15.30 $\pm$ 0.08  &  13.89 $\pm$ 0.07 &    --       &     --      &      --      &      --     &     2,4,8,X \\
 5 &    ...      &     --      &  15.98 0.10 &  12.65 $\pm$ 0.10 &  11.98 $\pm$ 0.11 &    --        &    --       &   \\
 8 &  19.72 $\pm$ 0.11 & 15.47 $\pm$ 0.02  &  13.15 $\pm$ 0.02 &  11.61 $\pm$ 0.16 &  11.32 $\pm$ 0.15 &    --        &    --       &   1,8,X \\
 9 &    --       &     --      &  15.55 $\pm$ 0.05 &    --       &     --      &     9.78 $\pm$ 0.12 &   7.25 $\pm$ 0.13 &   X    \\
 10 &    --       &  17.86 $\pm$ 0.12 &  16.24 $\pm$ 0.08 &    --       &     --      &     --        &    --        &    \\
11 &  15.15 $\pm$ 0.02 &  13.24 $\pm$ 0.02 &  12.36 $\pm$ 0.03 &    --       &     --      &      --       &     --       &   2,4,8,X \\
12 &    --       &  17.44 $\pm$ 0.12 &  14.27 $\pm$ 0.05 &  12.47 $\pm$ 0.18 &    --       &     --        &    --        &  1,9,X \\
13 &  16.36 $\pm$ 0.10 &  12.34 $\pm$ 0.08 &  10.14 $\pm$ 0.09 &   8.53 $\pm$ 0.04 &   7.87 $\pm$ 0.05 &   7.52 $\pm$ 0.04   & 6.85 $\pm$ 0.16  & 1,9,6,X \\
14 &  17.48 $\pm$ 0.08 &  15.56 $\pm$ 0.04 &  13.58 $\pm$ 0.03 &  11.37 $\pm$ 0.10 &  10.21 $\pm$ 0.10 &    --         &   --       &   2,9,4,5 \\
15   & 11.34 $\pm$ 0.03  &  9.79 $\pm$ 0.03   &  8.93 $\pm$ 0.03   &  8.06 $\pm$ 0.12   &  7.54 $\pm$ 0.11  &   6.78 $\pm$ 0.07  &  99.99 $\pm$ 9.99 &   
1,4,5,6,X \\ 
\hline
 \multicolumn{9}{l}{Remaining sources}\\ 
\hline
{ 25} &  11.93 $\pm$ 0.02 &   9.97 $\pm$ 0.03 &   9.18 $\pm$ 0.02 &  8.12 $\pm$ 0.03 &   7.84 $\pm$ 0.01  &  7.37 $\pm$ 0.03  &  6.77 $\pm$ 0.06  & 1,8,4 \\
{ 27} &  16.49 $\pm$ 0.01 &  14.18 $\pm$ 0.02 &  13.00 $\pm$ 0.01 &  11.96 $\pm$ 0.05 &  12.13 $\pm$ 0.05 &  10.61 $\pm$ 0.09 &   8.73 $\pm$ 0.13 &  2,4,X   \\
{ 35} &    --       &     --      &  16.60 $\pm$ 9.15 &  13.14 $\pm$ 0.03 &  12.00 $\pm$ 0.03 &    --       &     --      &    5  \\
{ 38} &  14.28 $\pm$ 0.02 &  12.78 $\pm$ 0.01 &  11.95 $\pm$ 0.01 &  11.05 $\pm$ 0.07 &  10.50 $\pm$ 0.10 &    --       &     --      &    2,8 \\
{ 39} &    --       &     --      &     --      &  14.07 $\pm$ 0.03 &  10.75 $\pm$ 0.03 &   8.89 $\pm$ 0.02 &   7.68 $\pm$ 0.03 &  5,7 \\
{ 42} &    --       &     --      &  14.40 $\pm$ 0.02 &  9.49 $\pm$ 0.03  &  7.29 $\pm$ 0.07  &  5.82 $\pm$ 0.06  &  4.46 $\pm$ 0.07 &  5,7,9, CHII-E\\
{ 43} &    --       &  16.48 $\pm$ 0.04 &  14.09 $\pm$ 0.04 &  9.25 0.08  &  7.27 $\pm$ 0.06  &  6.08 $\pm$ 0.04  &  5.08 $\pm$ 0.05 &  4,5,7 \\
{ 44} &  15.98 $\pm$ 0.01 &  13.53 $\pm$ 0.01 &  12.40 $\pm$ 0.02 &  10.93 $\pm$ 0.07 &  11.06 $\pm$ 0.05 &  10.09 $\pm$ 0.08 &    --      &    2,8,5,X \\
{ 45} &    --       &     --      &     --      &   13.22 $\pm$ 0.08 &  12.02 $\pm$ 0.03 &  10.99 $\pm$ 0.08 &    --     &     5,7,X\\
{ 49} &    --       &     --      &     --      &   13.28 $\pm$ 0.03 &  12.28 $\pm$ 0.09 &    --       &     --    &       X          \\      
{ 51} &  16.95 $\pm$ 0.02 &  13.83 $\pm$ 0.03 &  12.20 $\pm$ 0.03 &  10.59 $\pm$ 0.03  & 10.05 $\pm$ 0.04  & 10.88 $\pm$ 0.19  &  8.65 $\pm$ 0.04 &  1,4,5,6,8,X        \\       
{ 52} &    --       &  17.49 $\pm$ 0.04 &  15.88 $\pm$ 0.05 &    --        &    --       &     --      &      --     &    \\
{ 53} &    --       &     --      &  16.09 0.05 &  11.80 $\pm$ 0.02  & 10.17 $\pm$ 0.02  &  9.29 $\pm$ 0.02  &  8.67 $\pm$ 0.02  & 5,7,X \\
{ 60} &  10.97 $\pm$ 0.03 &  10.45 $\pm$ 0.03 &  10.10 $\pm$ 0.02 &   9.74 $\pm$ 0.07  &  9.68 $\pm$ 0.06  &  9.46 $\pm$ 0.07  &   --        &  3 \\
{ 65} &    --       &     --      &     --      &   15.59 $\pm$ 0.03 &  12.17 $\pm$ 0.07 &  12.01 $\pm$ 0.02 &   9.12 $\pm$ 0.15 &  5  \\
{ 75} &  19.40 $\pm$ 0.07 &  16.21 $\pm$ 0.02 &  14.67 $\pm$ 0.02 &  13.58 $\pm$ 0.02  & 13.44 $\pm$ 0.02  &   --        &    --       &   3,8 \\
\hline\hline
\label{tab:IR}
\end{tabular}
\end{center}
\tablefoot{1 = early B spectral type, 2 = late B to early A spectral type,
 3 = late-type, 4 = near-IR excess, 5 = mid-IR excess, 6 = Class II, 7 = Class~I,
 8 = 10 $< A_{V} <$ 28, 9 = $A_{V} >$ 28, and X = X-ray source (Feigelson et al. 2009).}

\end{table*}
 
\begin{table*}[!th]
\footnotesize
\renewcommand{\arraystretch}{1.5}
  \begin{center}
  \caption{Infrared photometry around the region of cometary nebulae. Units of the IR values are magnitudes.}
 \begin{tabular}{ccccccccccr}\hline\hline
 R.A.   & Dec. &      &            &             &          &        &       &       &       &    \\
 (J2000)  & (J2000)&      J&           H &           K  &          [3.6]&        [4.5] &       [5.8]  &       [8]&      { Notes*} &    Other\\
\hline
 \multicolumn{11}{l}{Around VLA~J172041.75$-$354808.2}\\ 
\hline
  17:20:41.92   &$-$35:48:04.7   &14.69$\pm$0.01   &13.40$\pm$0.01   &12.37$\pm$0.02     &9.50$\pm$0.15    &9.14$\pm$0.12   &6.46$\pm$0.12   &5.33$\pm$0.12  &2,8,4,5,7  &Mir-1 \\ 
  17:20:41.93   &$-$35:48:09.7   &  -          & -           & -             &10.52$\pm$0.18   &9.39$\pm$0.13   &6.64$\pm$0.13   &5.35$\pm$0.12  &5,7,X      &Mir-2  \\ 
 17:20:42.31   &$-$35:48:18.3   &21.35$\pm$0.13   &17.67$\pm$0.01   &14.40$\pm$0.02     &10.63$\pm$0.17   &9.52$\pm$0.12   &8.21$\pm$0.20    &-         &1,9,4,5,7  &Mir-3 \\
 17:20:42.19   &$-$35:48:12.5   &19.95$\pm$0.09   &17.12$\pm$0.03   &15.22$\pm$0.01       &-           &-           &-           &-         &2,9,4 &-   \\     
 17:20:42.00   &$-$35:48:07.7   &18.78$\pm$0.12   &16.45$\pm$0.04   &15.73$\pm$0.03       &-           &-           &-           &-         &3 &-       \\      
 17:20:42.02   &$-$35:48:11.0   &  -         &17.95$\pm$0.05   &15.80$\pm$0.09       &-           &-           &-           &-         &2,9  &-    \\     
 17:20:42.49   &$-$35:48:09.5   &21.48$\pm$0.12   &19.49$\pm$0.24   &16.56$\pm$0.04       &-           &-           &-           &-         &2,4 &-      \\     
 17:20:42.49   &$-$35:48:08.5   &  -          & -           &17.2$\pm$0.2        &-           &-           &-           &-       &-             \\    
 17:20:42.48   &$-$35:48:03.8   &  -          &-            &16.9$\pm$0.2        &-           &-           &-           &-       &-&-             \\      
 17:20:42.29   &$-$35:48:03.2   &20.01$\pm$0.10   &17.93$\pm$0.11   &16.79$\pm$0.04       &-           &-           &-           &-         &3&-          \\ 
 17:20:42.20   &$-$35:48:02.6   &20.33$\pm$0.17   &17.49$\pm$0.11   &16.52$\pm$0.06       &-           &-           &-           &-         &3  &-        \\ 
 17:20:42.02   &$-$35:49:10.1   &  -          &17.42$\pm$0.03   &16.70$\pm$0.07       &-           &-           &-           &-   &-  &-                \\
 17:20:42.15   &$-$35:48:17.8   &  -          &  -          &17.0$\pm$0.2        &-           &-           &-           &-   &-   &-                \\  
 17:20:42.11   &$-$35:48:13.2   &  -          &  -          &16.05$\pm$0.10       &-           &-           &-           &-   &-   &-                \\ 
 17:20:42.10  &$-$35:48:05.2    & -           & -           &15.8$\pm$0.2        &-           &-           &-           &-     &-   &-               \\ 
 \hline
 \multicolumn{11}{l}{Around VLA~J172041.59$-$354837.3}\\ 
\hline
17:20:41.58   &$-$35:48:36.9    &-  & 16.09$\pm$0.15  & 14.85$\pm$0.05  & 11.07$\pm$0.09  & 10.08$\pm$0.13 &   
7.14$\pm$0.10 &  5.51$\pm$0.15 & 7,X  &Mir-4 \\ 
\hline\hline
\label{tab:IRcome}
\end{tabular}
\end{center}
\tablefoot{{* Same as Table \ref{tab:IR}.}} 
\end{table*}
 
\section{Discussion}

To define the nature of the radio emission of the compact sources, 
we need information on the different characteristics ($\alpha$, polarization,
time variability). However we are only able to do a rough measure of the 
spectral index, which only gives clues on the emission mechanism. While 
thermal free-free radio emission has spectral indices with values 
$-0.1\leq\alpha\leq+2.0$, gyrosynchrotron emission has values in the 
range $-2.0\leq\alpha\leq+2.0$, and optically-thin synchrotron emission 
from wind collision regions has values  $\alpha\sim-0.7$ (see 
Rodr\'{\i}guez et al. 2012 and references therein). 

\subsection{Region D}
\subsubsection{Compact sources}
Although the reported compact radio sources are spread over the 
entire observed area, there is a group of 15 objects, reported here 
for the first time, that is well concentrated (within a radius of 0.3 pc) 
in the dark cloud on the western edge 
of region D. Of these, 73\% (11 sources) have unresolved infrared 
counterparts (see Fig.~\ref{fig:D}) and all but 3 are X-ray emitters 
(Feigelson et al. 2009). In the first parts of Tables~\ref{tab:RS} and ~\ref{tab:IR} 
we list the observational data of 
the compact sources in region D that we now focus on.

The clustering of compact radio sources clearly coincides with the 
densest part of the dark cloud, where one would suspect to have the
latest stage of star formation in the region. From the near-IR 
colors, we can distinguish only four stars that have a late-O or early-B spectral types,
while the rest are less massive. 
Two of these three massive stars have a negative spectral index, and thus their radio emission may originate in wind collision regions.
The majority of the remaining sources (eight sources) have negative spectral indices (non-thermal)  
values and only three are positive. The association of these sources with X-ray emission is compatible with a non-thermal origin for the radio emission, but there are a few cases where the spectral index is positive in spite of the presence of an X-ray source (Dzib~et~al.~2013).
As the sources are compact and  likely associated with the star forming region, an attractive interpretation is that they are magnetically
active low mass YSOs with gyrosynchrotron radio emission. However, the errors are 
large and this will have to be confirmed with multi-epoch monitoring to measure 
their flux variability,
as nonthermal low mass stars tend to be very variable on scales of days (Andr\'e 1996). 
These future observations will also help to obtain a better estimate of their spectral indices. 

It is worth highlighting that in the HII region associated with the Orion core 
{ (at d$\simeq$400 pc; Menten et al. 2007; Kounkel et al. 2017)}, there is a 
significant population of YSOs that produce radio emission  (e.g. Forbrich et 
al. 2016, Zapata et al. 2004). In order to compare with region D, if the Orion 
core were located 
at 1.34 kpc, the distance of the NGC~6334 complex, the flux density of the 556 compact
sources detected in Forbrich at al. (2016) would be 11.2 times lower. Considering 
50 $\mu$Jy as the detection threshold, only 47 compact sources from Forbrich et 
al. (2016) could be detected. We are only detecting 14 compact sources, or about one third of the expectation based on the Orion population. This is consistent with other similar comparisons (e.g., Masqu\'e 
et al. 2017), which may indicate that the Orion core may be richer in radio 
sources than other similar regions. Also, our images are highly contaminated by the
extended emission making it difficult to identify all compact sources.
Finally, there may be more sources below our detection threshold but 
we need deeper radio observations to clarify this point. 

\subsubsection{Cometary Nebula North (CNN) = VLA~J172041.75-354808.2}

Close to the north-western edge of the expanding HII region D, we discovered 
an extended radio source with a cometary  structure. From now on, we will 
refer to this source as CNN (for Cometary Nebula North). Its flux density 
at 6.0 GHz is 18.4$\pm$0.1 mJy with a peak flux of 
385$\pm12\, \mu$Jy bm$^{-1}$. The position of its peak flux is 
R.A.=\rahms{17}{20}{41}{75}; Dec.= \decdms{-35}{48}{08}{2}.  

This peak of the extended radio emission reported here is at the center of 
a mid-IR nebulosity with an ovoid shape of size around  $15''$. It is 
relatively bright in the 4.5, 5.8 and 8 $\mu$m {\it Spitzer}/IRAC images,
as shown in Fig. ~\ref{fig:CN}--Top. Although their spatial resolution 
(1\rlap{$''$}\,.5 to 
2\rlap{$''$}\,.0) is much worse than that obtained with the VLA,
a similar morphology of the diffuse emission is evident. The mid-IR structure 
is complicated by the presence of at least three bright, unresolved mid-IR 
sources (Mir-1, Mir-2 and Mir-3), most likely of protostellar nature embedded 
in the nebulosity. Fig. \ref{fig:CN}--Bottom shows the same field at 2.2 $\mu$m as observed in
excellent seeing conditions ($\sim 0\rlap{$''$}\,.55$) with the 2.5-m DuPont 
telescope at Las Campanas Observatory (Tapia, Persi \& Roth, in preparation). 
In addition to the mid-IR sources, there are a dozen near-IR stars detected within
the nebula. 
Comparison with the surrounding field shows that at least 60\% of them must be embedded 
in it. IRAC and $JHK$ photometry of these sources are reported in 
Table~\ref{tab:IRcome} and their colors are plotted in Figs. 3, 4, 5 and 6.

\begin{figure*}[!th]
   \centering
   \begin{tabular}{ccc}
  \includegraphics[height=0.36\textwidth,trim= 10 10 0 20, clip]{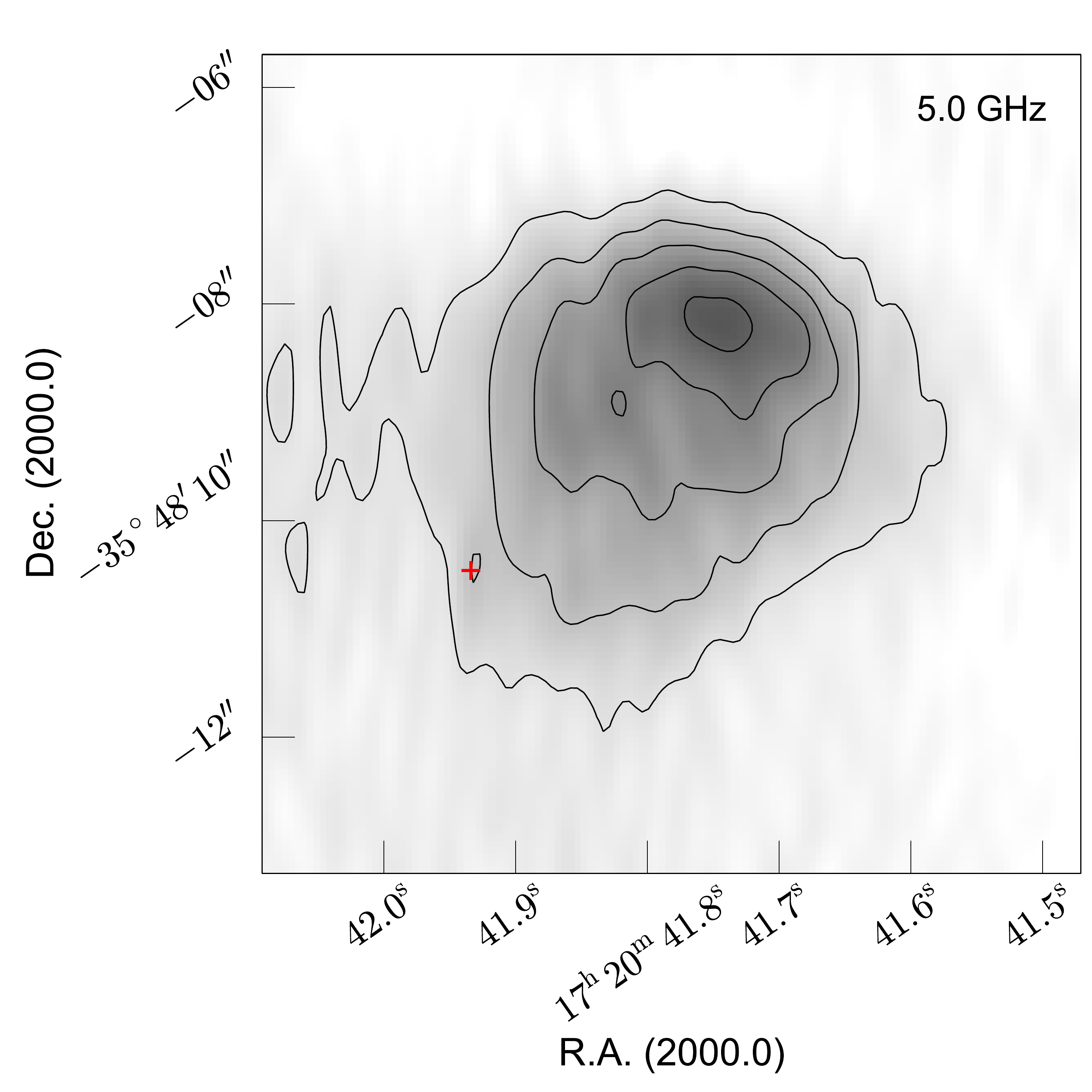} 
  & \includegraphics[height=0.36\textwidth,trim= 190 10 0 20, clip]{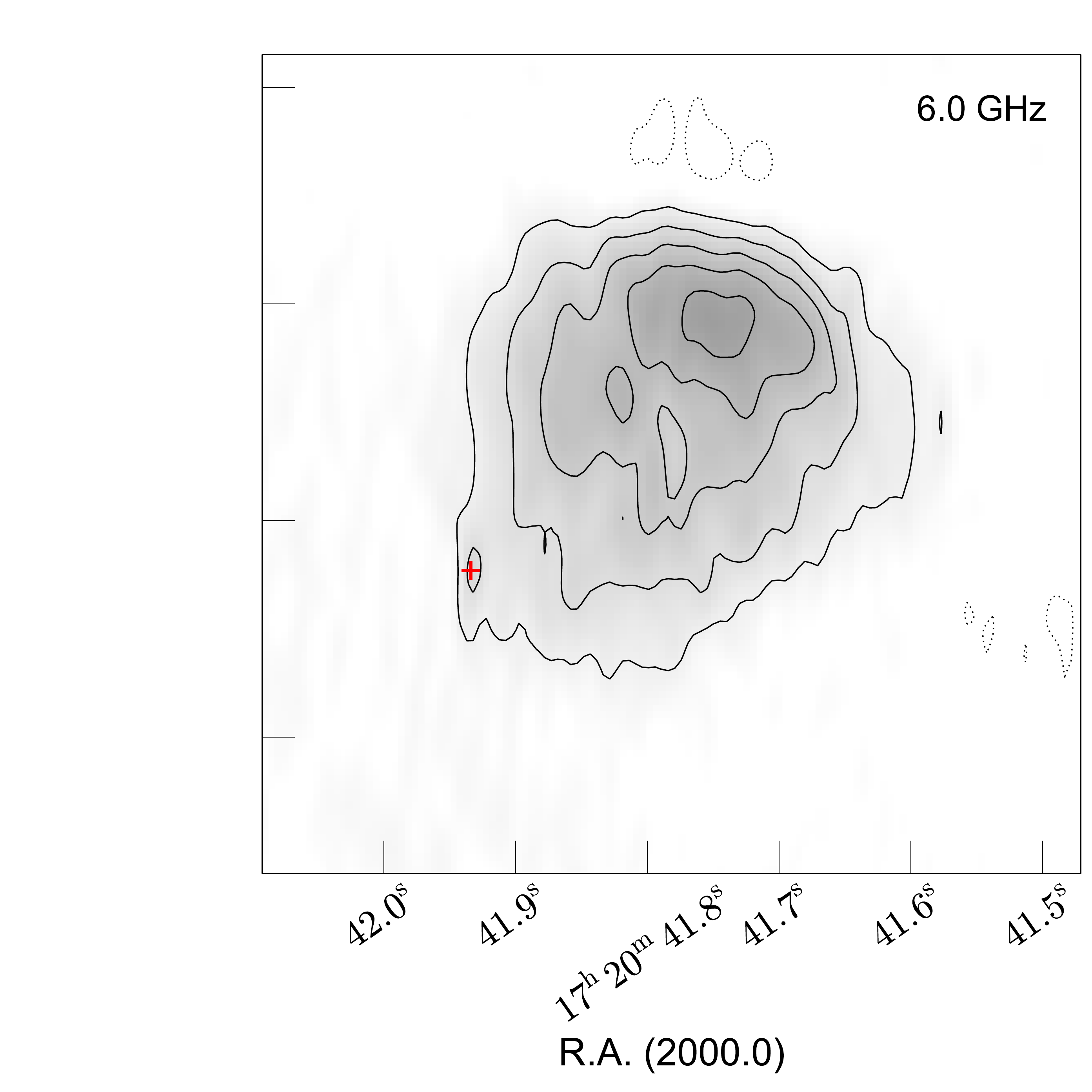}
  & \includegraphics[height=0.36\textwidth,trim= 190 10 0 20, clip]{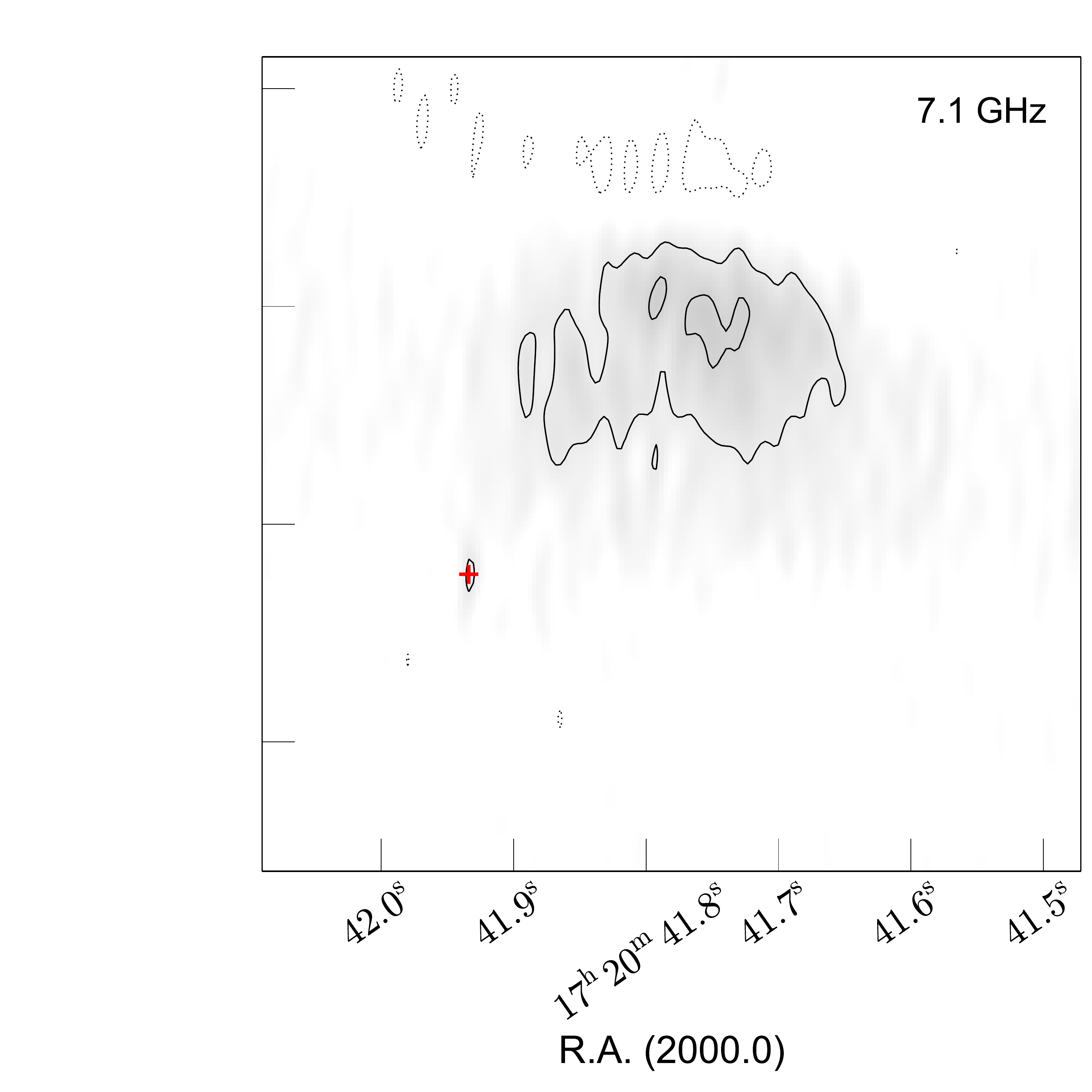}\\
  \end{tabular}
   \caption{Radio emission at the different frequencies from the cometary source close to region D.
   The noise levels around the source at each frequency are 25~$\mu$Jy, 15~$\mu$Jy and 19~$\mu$Jy
   for the 5.0~GHz, 6.0~GHz and 7.1~GHz maps, respectively. The contour levels are -3, 5,10, 15, 20 and 25 times the 
   noise level on each map. The red cross indicates the position of a source that is present in the three maps and may be related to Mir-2.}
   \label{fig:CNRadio}
\end{figure*}

The nature of the IR sources in CNN can be deduced from 
the multi-wavelength photometry. From the nebula-subtracted fluxes of sources 
Mir-1, Mir-2 and Mir-3, we conclude that they are Class I young stellar objects 
with spectral types earlier than A and each with quite different characteristics.
The notes in { column 9 of}  Table \ref{tab:IRcome} summarize the results. Interestingly, 
in the cometary nebulae there is another {emission peak} evident in the radio maps at 
the three frequencies (see Fig.~\ref{fig:CNRadio}) at the position 
R.A.=\rahms{17}{20}{41}{93}; Dec.= \decdms{-35}{48}{10}{5}. This position is 
consistent, within errors,
with the position of source Mir-2 and we suggest that it may be its radio 
counterpart.

Concerning the diffuse, nebular emission, from the present images, we attempted 
to measure nebular fluxes at several IR wavelengths. Because of many limitations 
(poor and very variable resolutions, presence of point-like sources, etc.) 
we found it impossible to deduce total fluxes reliably. Nevertheless, { by
integrating the near- and mid-IR flux densities that are well-derived from the 
ground-based and {\it Spitzer} observations and extrapolate them into the far-IR, 
we can estimate a lower limit to the total IR luminosity of the nebulosity to 
be  $\sim 10^3 L_\odot$.} Undoubtedly, this arises from thermal emission of warm dust. 

Thus we will assume that the CNN source is an \hii\ region, surrounded by a dusty envelope, 
and  derive its properties under that assumption. { At a frequency of 6.0~GHz,} we 
derive\footnote{ We have used the basic equations of brightness temperature, emission measure,
electron density, and ionizing photon rate (e.g., Table 4 in Masqu\'e et al. 2017).}
a brightness temperature of $44$~K, an electron density of 4.8$\times10^3$~cm$^{-3}$, an
emission measure of $0.6\times10^6$~cm$^{-6}$~pc, an ionized mass of 
8$\times10^{-3}$~M$_{\odot}$, and a flux of ionizing photons of 1.4$\times10^{46}$~s$^{-1}$
assuming a circular shape with a radius 
of 2$''$ = 0.013 pc, and an electron temperature of T$_e = 10^4$ K.
These values are consistent with an UC~\hii\ region being photoionized by a B0.5 ZAMS star
(Kurtz et al. 1994, Panagia 1973).  This is consistent with the approximated spectral 
types of sources Mir-1, Mir-2 and Mir-3.

\subsubsection{Cometary Nebula South (CNS) = VLA~J172041.59-354837.2}

As can be seen in Fig. \ref{fig:D}, a second small mid-IR nebula was 
found about $27''$ south of CNN (hereafter, we will refer to this source as 
CNS for Cometary Nebula South). Both nebulosities have similar mid- and far-IR 
colors implying matching dust
temperatures, though the integrated Herschel flux at 70 $\mu$m for CNS is four times fainter
than that from CNN. Unfortunately, no photometry could be obtained at longer
wavelengths and, thus, no reliable dust temperature can be derived. CNS has at its 
center a compact near- and
mid-IR source (Mir-4 in Fig. 9 and Table 3). The IR photometry of this source suggests that this
is an intermediate luminosity
Class I YSO with $A_V \simeq 13-16$. This object was also detected in X-rays by Feigelson et al.
(2009). At the same position, our 6.0 GHz image shows the presence of a small, roundish radio
source with a diameter of around $2''$. Its total flux { density} at 6.0~GHz is 4.5$\pm$0.4~mJy. As for CNN,
the presence of warm dust also indicates that the radio emission of CNS also traces ionized gas.
The derived parameter in this case are a
brightness temperature of $39$~K, an electron density of 6.2$\times10^3$~cm$^{-3}$, an
emission measure of $0.5\times10^6$~cm$^{-6}$~pc, an ionized mass of 
1$\times10^{-3}$~M$_{\odot}$, and a flux of ionizing photons of 3.0$\times10^{45}$~s$^{-1}$, which suggests that CNS is being photoionized by a B1 star (Panagia 1973). 

\begin{figure*}[!th]
   \centering
  \includegraphics[height=0.70\textwidth,trim= 20 20 50 50, clip]{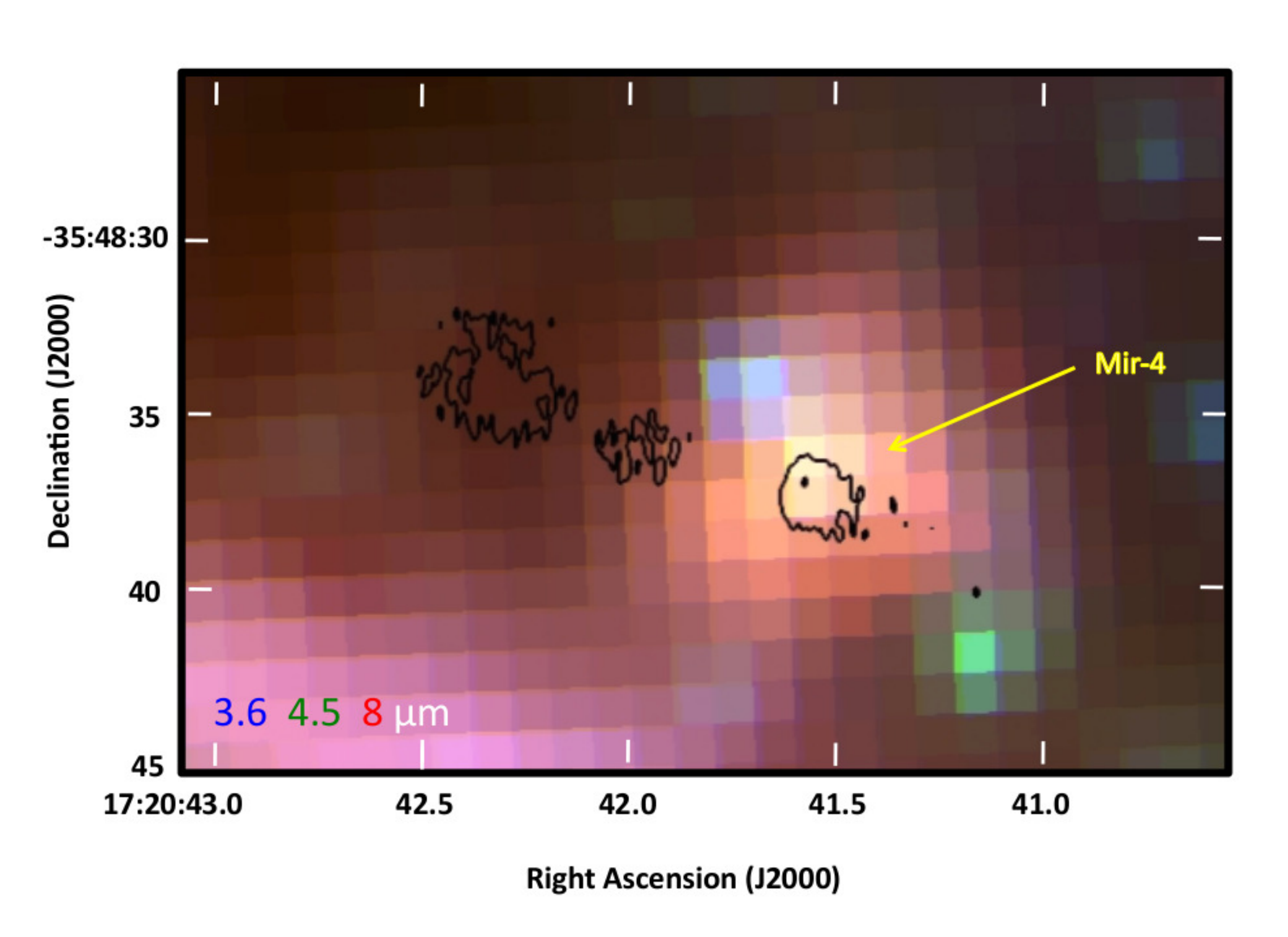}
   \caption{Composite infrared image around VLA~J172041.59$-$354837.3 (CNS).  Contours  correspond to the radio image at 6.0 GHz, and contour levels are the same as Fig.~\ref{fig:D}. { The yellow arrow indicate the positions of Mir-4 source.}}
   \label{fig:CNS}
\end{figure*}

It should be noted that the neighboring bright star that appears ``blue'' in the mid-IR IRAC image
(Fig.~\ref{fig:CNS}) is, given its IR colors, a foreground star unrelated to CNS.

\subsubsection{Double Nebula Source (DNS) = VLA~J172044.4-354917}

\begin{figure*}[!th]
   \centering
  \includegraphics[height=0.70\textwidth,trim= 20 20 50 30, clip]{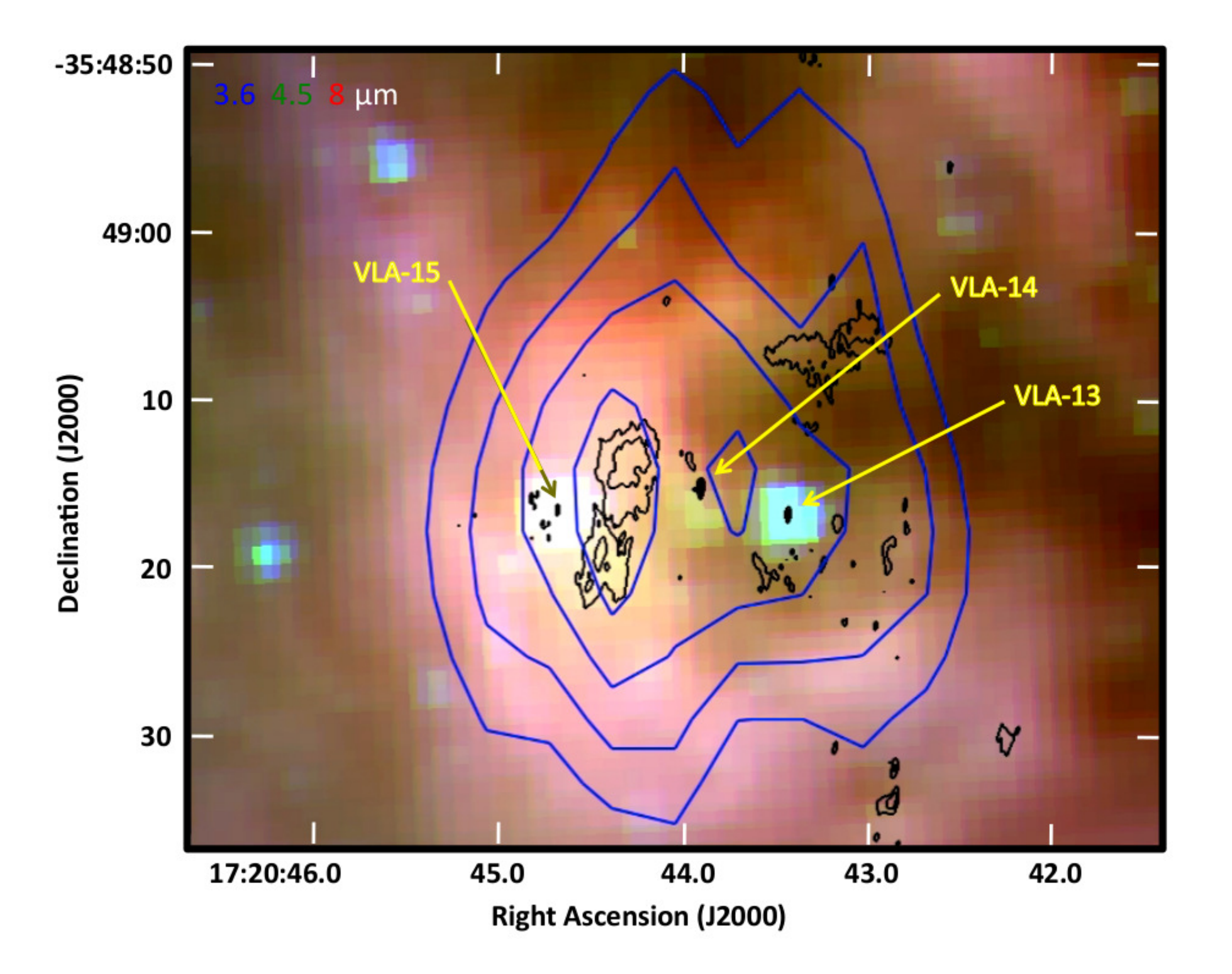}
   \caption{Composite infrared image around VLA~J1720444.4$-$354917 (DNS). Black contours  correspond to the
   radio image at 6.0 GHz, and contour levels are the same as Fig.~\ref{fig:D}. Blue
   contours correspond to the image 1.92 GHz, and the contour levels are 0.32, 0.37, 0.41 and 0.46 Jy/beam. 
   { The beamsize of this image is 26\rlap{.}$''$3$\times$6\rlap{.}$''$4; PA=$-179^{\circ}$.}}
   \label{fig:DNS}
\end{figure*}

Most intriguing is the extended double radio source VLA~J172044.47-3549017 
(hereafter DNS for Double Nebula Source), 
located almost at the center of the radio H II region D (e.g. the 1.6 GHz
map by Brooks \& Whiteoak 2001). This source is surrounded by three radio
compact sources (VLA~13, VLA~14 and VLA~15) with IR counterparts, see 
Fig.~\ref{fig:DNS}. Their IR counterparts indicate that they are massive 
stars with spectral types from late to early-B.

The nature of the double radio source is unknown. As can be seen in 
Fig.~\ref{fig:full} and Fig.~\ref{fig:DNS}, there is a lot of diffuse emission
in region~D. Thus, we speculate that DNS is gas ionized by the above mentioned 
three massive stars. 

To test this hypothesis we analyze additional VLA archive data.
The observations were made on 2013 July 25 in L (1 to 2 GHz) band under 
project 13A-448. The array was in the C configuration. The data were calibrated 
following the standard CASA procedures. To obtain the best angular resolution 
possible an image was made using the spectral window with the highest frequency,
centered at 1.92 GHz and with a bandwidth of 128 MHz. We additionally used
superuniform weighting in the CLEAN task of CASA. The image was also corrected
for the response of the primary beam and is shown in blue contours in Fig.~\ref{fig:DNS} . This image  indicates that source D has
a shell-like morphology and that the double radio source coincides with the 
brightest, eastern edge of the shell. 
Then, the double source may not be a true independent source but simply the 
brightest part of a larger structure.

\subsection{The central source in the C-$\hii$  region E}

\begin{figure*}[!th]
   \centering
  \includegraphics[height=0.70\textwidth,trim= 20 20 50 50, clip]{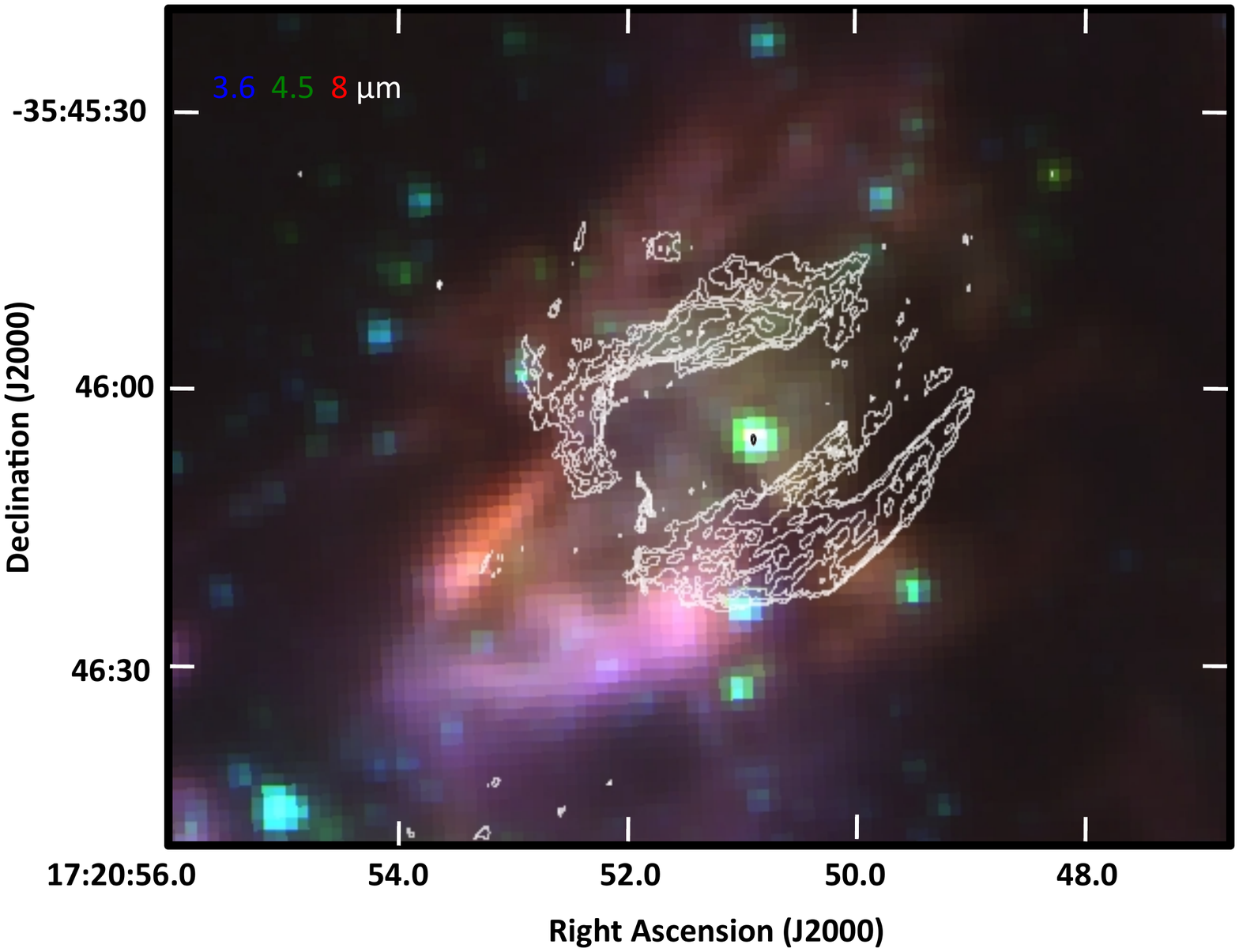}
   \caption{Composite infrared image around VLA~J172050.91$-$354605.0 (the source
   near the center of region E). Contours correspond to the
    radio image at 6.0 GHz, and contour levels are the same as Fig.~\ref{fig:D}. }
   \label{fig:E}
\end{figure*}

This compact radio source was discovered by Carral et al. (2002) close to the
center of the C-\hii\ region E (see Fig.~\ref{fig:E}) and it is our source 
No. 42  (Tables \ref{tab:RS} and 
\ref{tab:IR}). It is  coincident with IR source 161 of Tapia et al (1996), and 
was undetected in Feigelson et al.’s (2009) X-ray survey. Carral et al. (2002) 
measured a spectral index of $\alpha=1.0\pm0.7$ which  is
compatible with an ionized stellar envelope and, thus, they interpreted 
this source as the ionizing star of source E. 
The IR photometry here reported confirms that this is a Class I YSO, with 
$A_V > 55$ and $L_{\rm IR} > 10$ $L_\odot$. Lacking mid- and far-IR fluxes,
we cannot rule out, nor confirm that this star is the ionizing star of the 
C-\hii\ region, though the spatial coincidence makes it very probable. 

The measured spectral index in our maps is $\alpha=1.1\pm0.3$ which is 
consistent with the previous results and the thermal nature of this source is
corroborated. Using the estimated spectral index to extrapolate our measured 
flux to 8.4 GHz, we obtain a flux density of 1.2$\pm$0.3 mJy,
which is in good agreement with the flux density measured by Carral et al. (2002).
This non-variability also supports the thermal nature of the radio source.

\subsection{Long term variability}

Most of the previous radio observations on NGC 6334 were focused on
regions E, F, and I(N), as they have shown the most recent star formation
on this molecular cloud. Furthermore, they reached noise levels 
that were much poorer than those of the final images reported on this paper.
Still, we may use these observations to roughly measure variability
on some sources. In the VLA archive we found a VLA observation, in A 
configuration, obtained on 11 August 1995 at 8.4 GHz, that is part of project 
AM495. These observations were reported in Carral et al. (2002). 
The pointing center was RA=\rahms{17}{20}{53}{40}; 
Dec.=\decdms{-35}{46}{25}{0} (i.e. roughly equidistant from
regions E, F, and I(N)). The data were edited and calibrated in a standard 
way. Images were produced  with a pixel size of 0\rlap{$''$}\,.06 and, as it was 
done by Carral et al. (2002), removing the short spacings below 100 k$\lambda$ 
(to filter out structures larger than 2$''$). { The final beamsize of this image was
0\rlap{.}$''$48$\times$0\rlap{.}$''$19; PA=0\rlap{$^\circ$}\,.7.}
We look at the position of sources
in Table \ref{tab:RS} and that are inside the imaged area. In this case, we
used a threshold of three times the noise level to consider a detection.
Five sources were detected using this restriction and their fluxes from
this image are reported in Table~\ref{tab:95vs2011}. Additionally, 
using the calculated spectral index, 
we extrapolate our measured fluxes at 6.0 GHz and determine their fluxes at 
8.4 GHz, these are also listed on Table \ref{tab:95vs2011}. Most of the sources
are in good agreement with their expected flux in 2011, which suggests that these
sources are not strongly variables. The only exception is source 
VLA~J172058.14$-$354934.6 which shows a decrease in its flux from 1995 to 2011 
by a factor of $2.4\pm0.3$.

On the other hand, we notice that due to the smaller imaged area and 
larger noise, most of the sources are not expected to be detected because they fall
outside the primary beam or their flux will be below three 
times the noise level. However, using the extrapolated flux in epoch 2011
and assuming non-variability, three sources should be detected above this 
threshold in the 1995 image. Interestingly, they are not. 
In Table \ref{tab:95vs2011} we list these sources, an upper limit of three 
times the noise level (with the noise level of the area as the flux error)
and its predicted flux at 8.4 GHz { for the epoch 2011}. Clearly sources VLA~J172052.02$-$354938.0
and VLA~J172053.65-354548.4 are strong variables { ($\gtrsim$100\%)} on scales of years. 
We could not support a variability for source VLA~J172057.98$-$354431.6, since its non-detection 
may be due only to fluctuation in the flux caused by the noise of the observations.

We also searched for possible additional new sources by using BLOBCAT 
and with the parameters used for the 6.0 GHz map. We did not find any
additional new source.

The three strongly variable sources reported here may be good examples of magnetically active 
YSOs. With future deep observations we could study variability from the weakest 
radio sources.

\begin{table}[!t]
\footnotesize
\renewcommand{\arraystretch}{1.51}
  \begin{center}
  \caption{Comparison from radio sources of epoch 1995 and 2011.}
 \begin{tabular}{crr}\hline\hline
                   & $S_\nu$ (1995) & $S_\nu$ (2011) \\ 
VLA name            & ($\mu$Jy) & ($\mu$Jy){\smallskip}\\
\hline
J172103.47$-$354618.8 &   406$\pm$59     &   453$\pm$52\\
J172052.02$-$354938.0 &  $<408\pm$136    &  1038$\pm$93\\
J172053.65$-$354548.4 &   $<171\pm$57    &    338$\pm$62\\
J172057.98$-$354431.6 &   $<216\pm$72    &    220$\pm$24\\
J172050.91$-$354605.0 &   $1380\pm$110   &   1188$\pm$290\\
J172055.19$-$354503.8 &    401$\pm$107   &    365$\pm$23\\
J172058.14$-$354934.6 &   1240$\pm$110   &    510$\pm$52\\
J172054.62$-$354508.5 &    290$\pm$60    &    192$\pm$25\\
\hline\hline
 \label{tab:95vs2011}
\end{tabular}
\end{center}
\tablefoot{Upper-limits to fluxes are obtained as three times the noise level,
that are at the same time used as the error for these upper-limits.}
\end{table}



\subsection{Other sources}

We could not determine the spectral classification for most of the remaining radio sources.
However, the infrared emission of two sources, 24 and 50 in Table \ref{tab:RS}, indicate that they are 
early B  stars. Their spectral index has large errors but suggest a flat spectrum, 
so they are most probably thermal emitters. The radio emission in these cases may   originate in the winds of the massive stars. 

It is hard to speculate 
on the radio emission of the remaining sources, because of the scarce information.
Future multi-wavelength and multi-epoch observations with better sensitivity may 
help to reveal their nature.

\section{Conclusions}

We have presented a deep radio observation ($\sigma\sim50\,\mu$Jy bm$^{-1}$) 
with high angular resolution (0\rlap{$''$}\,.2)
of the NE of the NGC 6334 complex (covering the regions D, E, F, I(N), and 
part of the region C) searching for compact radio sources. 
We also searched for infrared counterparts of detected compact radio sources 
and characterized them. Now we list the results and conclusions from our analysis.

\begin{itemize}

\item A total of 83 radio compact sources in the NE of NGC 6334 are detected, 
15 of them are located inside the region D. 
Most of these sources are new detections and only around 10
of them were previously reported and are located in regions E, I, and I(N).

\item The stellar nature of 27 of the 83 compact radio sources is confirmed by
the properties of their infrared emission. 

\item We computed the spectral index of the sources in order to 
speculate about the  nature of their radio emission. In region~D the values 
tend to be negative, suggesting non-thermal emission. Most of these 
sources are likely magnetically active low mass YSOs as in the Orion core. 
However, the IR emission of three of them suggest that they are early B 
stars and their radio emission may originate in strong shocks of wind 
collision regions.

\item Two interesting cometary radio sources, CNN and CNS, were detected close to region~D and
are here reported  for the first time. They are spatially coincident with more extended mid-IR 
nebulosities of similar shape. We suggest that they are \hii\ regions (traced
by the radio emission) surrounded by dusty envelopes (traced by the mid-IR).
Interestingly we found three stars (Mir-1, Mir-2 and Mir-3) with spectral types
earlier than A, which could be the sources of the ionizing photons of CNN. On the other hand, 
the possible ionizing source of CNS is Mir-4. 

\item  Through the inspection of an additional 1.92 GHz image, we suggest that the double 
source VLA~J1720444.4-354917 (DNS) is part of the  diffuse ionized gas from the region D.

\item Our observations support the thermal nature of source VLA J172050.91$-$354605.0,
which is located near the center of the radio \hii\ region E. 

\item By comparing with an observation obtained in 1995, we analyzed the variability 
in flux of eight sources. Three of them show strong variability  { suggesting} that
they are magnetically active low mass stars.


\end{itemize}

Our analysis has provided  clues on the nature of several of the detected 
compact radio sources. However, future observations are necessary to better 
establish the nature of most of them. Ideally, these observation should be 
multi-wavelength and multi-epoch.

\begin{acknowledgements}
S.-N.X.M. acknowledges IMPRS for a Ph.D. research scholarship.
L.F.R., M. T. and L.L. acknowledge the financial 
support of DGAPA, UNAM, and CONACyT, M\'exico.
MT and LL acknowledge support for this work through 
UNAM-PAPIIT grants IN104316, and IN112417. This paper makes
use of archival data obtained with
the {\sl Spitzer Space Telescope}, which is operated by the Jet Propulsion Laboratory, California
Institute of Technology (CIT)
under National Aeronautics and Space Administration (NASA) contract 1407.
This research has made use of the NASA/ IPAC Infrared Science Archive, which is operated by
the Jet Propulsion Laboratory, California Institute of Technology, under contract with the National
Aeronautics and Space Administration

This research has made use of
the SIMBAD database, operated at CDS, Strasbourg, France.
\end{acknowledgements}

\bibliographystyle{aa}

\clearpage
\onecolumn

\end{document}